\PassOptionsToPackage{table}{xcolor}
\documentclass{article} % For LaTeX2e
\usepackage{arxiv,times}

\usepackage{amsmath,amsfonts,bm}

\def\eqref#1{equation~\ref{#1}}
\def\1{\bm{1}}

\def\rvc{{\mathbf{c}}}

\def\rvx{{\mathbf{x}}}

\def\vs{{\bm{s}}}

\def\vu{{\bm{u}}}
\def\vv{{\bm{v}}}

\def\vy{{\bm{y}}}

\DeclareMathAlphabet{\mathsfit}{\encodingdefault}{\sfdefault}{m}{sl}
\SetMathAlphabet{\mathsfit}{bold}{\encodingdefault}{\sfdefault}{bx}{n}

\usepackage{float}
\usepackage{graphicx}
\usepackage{hyperref}
\usepackage{url}
\usepackage{booktabs}
\usepackage{multirow}
\usepackage{tabularx}
\usepackage{threeparttable}
\usepackage[most]{tcolorbox}
\usepackage{makecell}
\usepackage{wrapfig}
\usepackage{capt-of}
\usepackage{algorithm}
\usepackage{algpseudocode}

\title{SEmoEdit: Probing and Harnessing the Editability of Pre-trained Speech Flows}

\author{Tianxin Xie, Pengfei Zhang, Kai Jiang, Zelin Zhao, Li Liu\footnote{Corresponding Author}}
\author{Tianxin Xie\textsuperscript{\rm 1},
Pengfei Zhang\textsuperscript{\rm 1},
Kai Jiang\textsuperscript{\rm 1},
Zelin Zhao\textsuperscript{\rm 1},
Li Liu\textsuperscript{\rm 1}\thanks{Corresponds to Li Liu (avrillliu@hkust-gz.edu.cn)} \\
\textsuperscript{\rm 1}The Hong Kong University of Science and Technology (Guangzhou)
}

\iclrfinalcopy % Uncomment for camera-ready version, but NOT for submission.
\begin{document}

\maketitle

\begin{abstract}
Existing training-based speech emotion editing methods often require substantial task-specific training and can be unstable.
This motivates us to investigate whether the pretrained generative dynamics of large-scale text-to-speech (TTS) models can be directly manipulated for training-free emotion editing.
To answer this question, we probe the editability of pretrained flow-matching and hybrid TTS models by constructing a controlled test set and systematically diagnosing editing effects along the generative trajectory.
Our analysis reveals that pretrained TTS models are substantially editable in emotion, but such editability is architecture- and trajectory-dependent and can be disrupted by early flow-matching steps, while cross-speaker emotion transport carries additional acoustic attributes beyond emotion.
To address these limitations, we propose \textbf{SEmoEdit}, the first training-free framework that formulates emotion editing as dynamic velocity transport between source and target emotions, enabling robust, flow-based speech emotion editing directly within pretrained TTS models.
SEmoEdit unifies three core operations: emotion replacement, emotion erasure, and continuous emotion interpolation, requiring neither parameter updates nor task-specific optimization.
To systematically evaluate these capabilities, we introduce \textbf{SEmoEditBench}, a dataset comprising 600 editing cases, and conduct extensive experiments across state-of-the-art (SOTA) models and backbones.
Our results show that SEmoEdit is highly effective and broadly applicable, outperforming existing training-based and activation-steering methods.
Ultimately, this work reveals that pretrained speech flows possess rich, latent emotion-editing capabilities, providing useful guidance for real applications.
Code, benchmark, and Audio samples are available at \url{https://github.com/imxtx/SEmoEdit}.
\end{abstract}

\section{Introduction}
\label{sec:introduction}

Flow-matching TTS formulates speech generation as transport along a learned velocity field, enabling efficient non-autoregressive synthesis \citep{guo2024voiceflow,eskimez2024e2,chen2025f5ttsfairytalerfakesfluent}. Large-scale hybrid TTS models further integrate autoregressive generation with flow matching, achieving strong zero-shot voice cloning and expressive synthesis capabilities \citep{du2024cosyvoice2scalablestreaming,zhou2026indextts2,hu2026qwen3}.
While controllable speech synthesis has been extensively studied \citep{xie-etal-2025-towards}, these advances raise a natural question beyond generation: \textit{can the capabilities encoded in pre-trained TTS models be directly harnessed to edit the emotion of an existing utterance while preserving its linguistic content and speaker identity?}
In this paper, we define three speech emotion editing tasks as proxies for exploring this question:
% We define three speech emotion editing tasks:
(1) \emph{emotion replacement}, replacing the source emotion with a target emotion; (2) \emph{emotion erasure}, neutralizing the source emotion; and (3) \emph{emotion interpolation}, continuously traversing between source and target emotions.
% All tasks edit only the affective expression of an existing utterance without model training, distinguishing training-free emotion editing from emotion-controllable TTS, which generates new utterances \citep{zhou2026indextts2}, and emotional voice conversion, which relies on task-specific conversion models \citep{ZHOU20221}.

Existing methods achieve related capabilities primarily through emotion conditioning or activation steering.
Conditional generation approaches such as PromptTTS \citep{PromptTTS}, EmoSphere++ \citep{Cho2025EmoSphere++}, and IndexTTS2 \citep{zhou2026indextts2} learn label-, prompt-, or reference-based representations to enable emotional speech generation.
Dedicated emotion editing methods, such as dots.tts.edit \citep{wang2026dotsedit} and Bagpiper-Edit \citep{gong2026bagpiper}, further support emotion modification by localizing and regenerating selected spans of the acoustic representation.
More recently, activation-steering-based methods, e.g., EmoSteer-TTS \citep{xie2025emosteer} and CoCoEmo \citep{wang2026cocoemo}, derive emotion directions from activation differences between neutral and emotional speech and apply vector arithmetic to control emotion without fine-tuning the base model.

However, these approaches face several limitations. (1) They often require substantial resources beyond the pre-trained model. Dedicated editing methods usually depend on large-scale edit-specific datasets for training, while methods such as EmoSteer-TTS rely on paired emotional corpora, auxiliary emotion recognition models, or model-specific probing to derive steering vectors. (2) Although training-free, activation-steering methods are largely static, applying fixed activation directions with global coefficients at selected layers, tokens, or sampling steps, leading to unstable or unsuccessful results. (3) They lack a unified mechanism across editing tasks: different objectives or model architectures often require distinct input/output/model designs, curated datasets, or task-specific training.

To address these challenges and achieve robust training-free emotion editing, we need to answer three questions: \textit{(1) Can pretrained speech flows be directly edited without training? (2) When along the generative trajectory should the editing be applied? (3) What acoustic attributes are actually carried by a source-to-target velocity transport?}
Our study yields three observations:
First, pretrained speech flows possess substantial direct editability.
Second, editability is strongly non-uniform along the trajectory.
Third, conditional speech velocity fields are not attribute-factorized.

Guided by these observations, we propose \textbf{SEmoEdit}, a training-free framework that recasts speech emotion editing as dynamic velocity transport between source and target emotions.
Instead of injecting a fixed global activation direction, SEmoEdit dynamically estimates the source-to-target velocity transport process throughout sampling.
The same formulation supports emotion replacement, erasure, and interpolation by controlling the editing strength of the emotional transport, without parameter updates or task-specific optimization.
To systematically evaluate our framework, we introduce \textbf{SEmoEditBench}, a benchmark containing 600 speech emotion editing cases.
We evaluate SEmoEdit on three different TTS backbones \citep{du2024cosyvoice2scalablestreaming,chen2025f5ttsfairytalerfakesfluent,zhou2026indextts2} and compare it directly with leading training-based \citep{yan2025stepaudioeditx,wang2026dotsedit} and activation-steering approaches \citep{xie2025emosteer,wang2026cocoemo}.
Our results consistently demonstrate highly effective emotion editing across diverse models and tasks, revealing that pretrained TTS systems harbor latent emotion-editing capabilities that can be unlocked via inference-time velocity control.
Our in-depth analysis of the editability also provides guidance for future research and applications.
The contributions of this paper are summarized as follows:

\begin{itemize}
    \item We present the first systematic study of pretrained speech flows' editability, asking and answering whether, when, and under what conditions an utterance can be directly edited without sacrificing intelligibility, naturalness, speaker identity, or emotional authenticity.
    \item We propose SEmoEdit, the first training-free framework to formulate speech emotion editing as dynamic velocity transport. By directly manipulating the velocity field, SEmoEdit seamlessly unifies emotion replacement, erasure, and continuous interpolation.
    \item We introduce SEmoEditBench and evaluate across SOTA models and backbones. Extensive comparisons demonstrate our framework's superiority over existing approaches, proving that robust emotion editing can be unlocked directly within pretrained TTS systems.
\end{itemize}

% \begin{table*}[t]
%     \centering
%     \caption{Comparison of representative speech emotion editing methods and our proposed SEmoEdit.}
%     \label{tab:method_capability_comparison}
%     \scriptsize
%     \setlength{\tabcolsep}{4pt}
%     \begin{tabularx}{\textwidth}{@{} l c c c c c >{\raggedright\arraybackslash}X @{}}
%         \toprule
%         Method & \makecell{Required \\ Data Scale} & Training-free & Replace & Erase & Interpolate & Mechanism \\
%         \midrule
%         Step-Audio-EditX \citep{yan2025stepaudioeditx} & Not reported & -- & $\checkmark$ & $\checkmark$ & -- & Instruction-guided regeneration \\
%         dots.tts.edit \citep{wang2026dotsedit} & 10M samples & -- & $\checkmark$ & $\checkmark$ & -- & Instruction-guided regeneration \\
%         % AuK \citep{ma2026auk} & Not reported & -- & $\checkmark$ & $\checkmark$ & -- & Instruction-guided regeneration \\
%         EmoSteer-TTS \citep{xie2025emosteer} & 9.6K samples & $\checkmark$ & $\checkmark$ & $\checkmark$ & $\checkmark$ & Fixed activation steering \\
%         CoCoEmo \citep{wang2026cocoemo} & 20K samples & $\checkmark$ & $\checkmark$ & -- & $\checkmark$ & Fixed activation steering \\
%         \textbf{SEmoEdit (Ours)} & 1 sample & $\checkmark$ & $\checkmark$ & $\checkmark$ & $\checkmark$ & Dynamic velocity transport \\
%         \bottomrule
%     \end{tabularx}
% \end{table*}

\section{Analyzing the Editability of Pre-trained Speech Flows}
\label{sec:analysis}

% This section addresses the \textit{``whether editable''}, \textit{``when to edit''}, and \textit{``what to transport''} questions proposed in Sec.~\ref{sec:introduction} by analyzing the editability of pre-trained flow-matching and hybrid TTS models.
% We first review conditional flow matching and formulate the emotion editing problem, then examine the editability of existing TTS models and identify the challenges that motivate our methods.

\subsection{Preliminaries}

\textbf{Conditional Flow Matching (CFM).}
Let $\rvx_1 \in \mathbb{R}^{D \times L}$ be a mel spectrogram and $\rvc$ its synthesis condition (e.g., text, speaker, paralinguistics). CFM \citep{lipman2023flow} learns a velocity field transporting a prior $p_0(\rvx)$ (Gaussian or semantic tokens) to the speech distribution $p_1(\rvx \mid \rvc)$. For data $\rvx_1 \sim p_1(\rvx \mid \rvc)$ and noise $\rvx_0 \sim p_0(\rvx)$, the optimal-transport path and conditional velocity are:
\begin{equation}
    \rvx_t = (1-t)\rvx_0 + t\rvx_1, \qquad \vu_t(\rvx_t \mid \rvx_1) = \rvx_1 - \rvx_0, \qquad t \in [0,1].
    \label{eq:cfm_path}
\end{equation}
A neural velocity field $\vv_\theta(\rvx_t,t;\rvc)$ is trained with $t \sim \mathcal{U}[0,1]$ via:
\begin{equation}
    \mathcal{L}_{\mathrm{CFM}}(\theta) = \mathbb{E}_{t,\rvx_1,\rvx_0}\!\left[\left\|\vv_\theta(\rvx_t,t;\rvc)-\vu_t(\rvx_t \mid \rvx_1)\right\|_2^2\right].
    \label{eq:cfm_objective}
\end{equation}
At inference, a sample is generated by solving the ODE from $\rvx(0) \sim p_0$:
\begin{equation}
    \frac{\mathrm{d}\rvx(t)}{\mathrm{d}t}=\vv_\theta(\rvx(t),t;\rvc), \qquad \rvx(1) \sim p_1(\rvx \mid \rvc).
    \label{eq:cfm_ode}
\end{equation}
This formulation underpins modern flow-matching \citep{mehta2024matcha,eskimez2024e2,chen2025f5ttsfairytalerfakesfluent} and hybrid TTS models \citep{du2024cosyvoice2scalablestreaming,hu2026qwen3}.

\textbf{The Emotion Editing Problem.}
We formulate ideal speech emotion editing as transforming a source utterance $\rvx^{\mathrm{src}} \sim p_1(\rvx\mid \vy,\vs,e_{\mathrm{src}})$, conditioned on linguistic content $\vy$, speaker $\vs$, and emotion $e_{\mathrm{src}}$, into a target emotion $e_{\mathrm{tgt}}$:
\begin{equation}
    \rvx^{\mathrm{edit}}
    =
    \mathcal{E}\!\left(
        \rvx^{\mathrm{src}};
        e_{\mathrm{src}}\!\rightarrow e_{\mathrm{tgt}}
    \right),
    \qquad
    \rvx^{\mathrm{edit}}
    \sim
    p_1(\rvx\mid \vy,\vs,e_{\mathrm{tgt}}).
    \label{eq:emotion_editing}
\end{equation}
A perfect edit preserves content $\mathcal{C}(\cdot)$ and speaker $\mathcal{S}(\cdot)$ while exclusively updating the emotion $\mathcal{A}(\cdot)$:
\begin{equation}
    \mathcal{C}(\rvx^{\mathrm{edit}}) = \mathcal{C}(\rvx^{\mathrm{src}}),
    \qquad
    \mathcal{S}(\rvx^{\mathrm{edit}}) = \mathcal{S}(\rvx^{\mathrm{src}}),
    \qquad
    \mathcal{A}(\rvx^{\mathrm{edit}}) = e_{\mathrm{tgt}}.
    \label{eq:emotion_edit_constraints}
\end{equation}
Since pre-trained TTS models learn generation under entangled conditions rather than factorized editing operators, we investigate \textit{whether} their learned velocity fields support such editing, \textit{when} the editing signal is effective along the trajectory, and \textit{what} attributes it actually carries.

\subsection{The Editability of Flow-Matching and Hybrid TTS Models}

\textbf{Q1: Can pre-trained TTS models be directly edited without additional training?}

To test if pre-trained velocity fields contain a usable editing signal, we probe frozen F5-TTS and CosyVoice 2 using 80 parallel neutral-to-emotion pairs (matched text and speaker) from ESD \citep{ZHOU20221}, covering 20 speakers and four emotions (Happy, Angry, Sad, Surprise).

\begin{figure}[t]
    \centering
    \includegraphics[width=\textwidth]{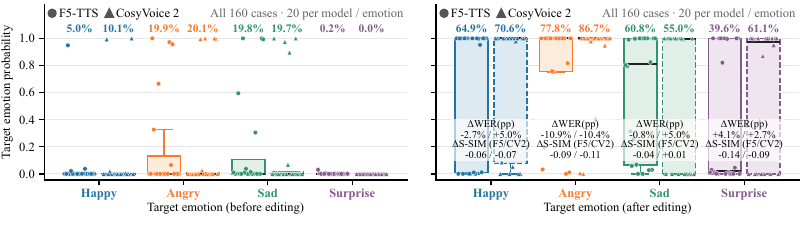}
    \caption{Target emotion2vec \citep{ma2024emotion2vec} probability, change in transcription error ($\Delta$WER, in percentage points; pp), and change in speaker similarity ($\Delta$S-SIM) before and after editing.}
    \label{fig:q1_target_probability}
\end{figure}

We synthesize a source mel \(\rvx^{\mathrm{src}}\), length-match it to the target condition, and construct a diagnostic probe inspired by inversion-free image editing \citep{kulikov2025flowedit,xu2024inversionfree}. The probe uses the conditional velocity difference under shared noise as a local editing signal:
\[
\Delta\vv_\theta(t)
=
\vv_\theta\!\left(\overline{\rvx}_t^{\mathrm{probe}},t;\rvc^{\mathrm{tgt}}\right)
-
\vv_\theta\!\left(\overline{\rvx}_t^{\mathrm{src}},t;\rvc^{\mathrm{src}}\right).
\]
Accumulating this signal along the flow trajectory effectively shifts \(\rvx^{\mathrm{src}}\) toward the target emotion (complete transport formulation in Sec.~\ref{sec:semoedit}).
As Fig.~\ref{fig:q1_target_probability} shows, this substantially increases target-emotion probability across all emotions, albeit with mixed Word Error Rate (WER) changes and moderate speaker similarity (S-SIM) degradation (details in Appendix~\ref{app:editability_diagnostic_q1}).

\begin{tcolorbox}[colback=violet!8!white,colframe=black,boxrule=0.7pt,arc=3pt,
  left=6pt,right=6pt,top=5pt,bottom=5pt,boxsep=0pt]
\noindent\textbf{Observation 1:}
Pre-trained speech velocity fields contain a directly usable source-to-target editing signal, enabling emotion modification without model training.
\end{tcolorbox}

\textbf{Q2: When along the generative trajectory should editing be applied?}

\begin{figure}[t]
    \centering
    \includegraphics[width=\textwidth]{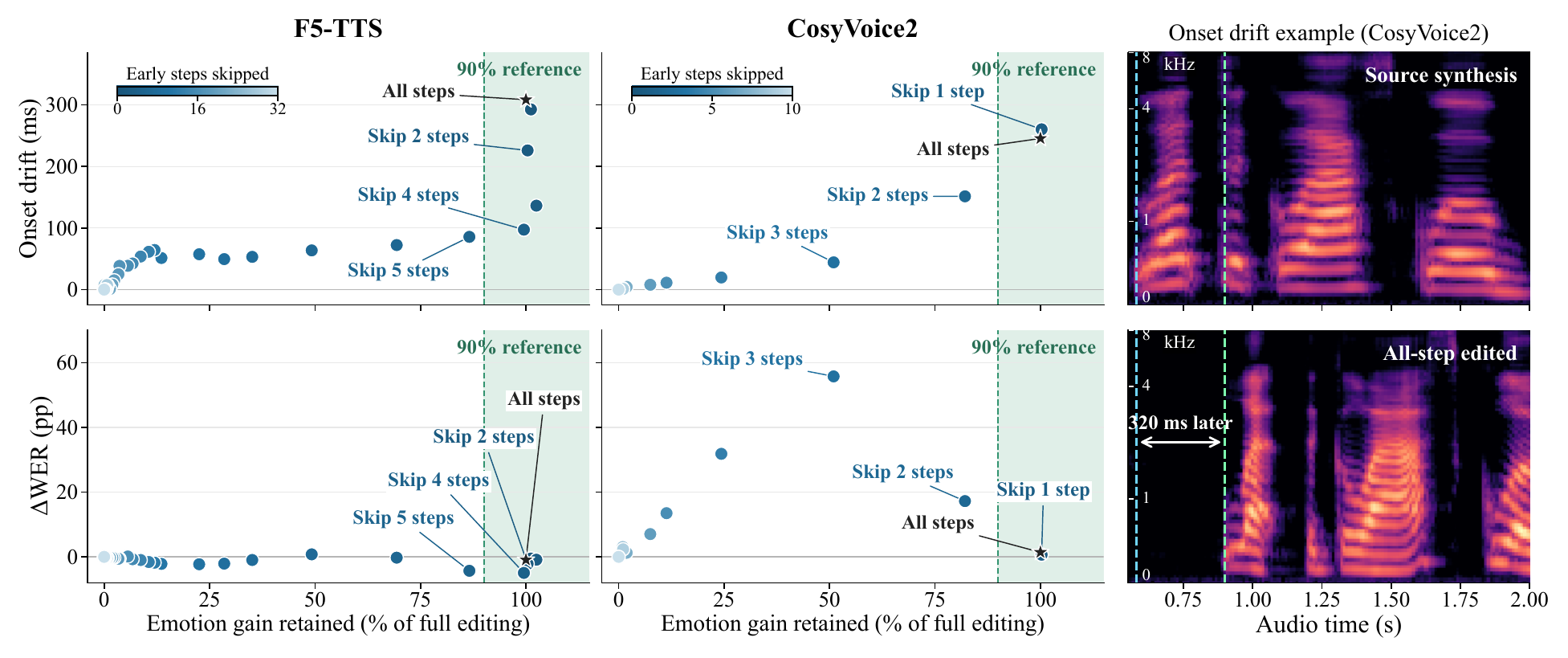}
    \caption{The first two columns show onset drift (top) and WER change from the source in percentage points (bottom) versus retained emotion gain relative to full-step editing. The green region denotes $\geq$90\% gain retention. The third column demonstrates the onset drift problem.}
    \label{fig:q2_editing_steps}
\end{figure}

Building on Q1, we examine whether the editing signal's effectiveness varies along the generative trajectory. Full-step transport successfully transfers emotion but introduces temporal distortions, notably \emph{\textbf{onset drift}} (mean 308 ms for F5-TTS, 245 ms for CosyVoice 2). As Fig.~\ref{fig:q2_editing_steps} shows, successful emotion transfer and zero Word Error Rate (WER) do not guarantee temporal preservation.

To locate where this distortion arises, we delay transport by $k$ integration steps and measure the retained emotion gain, onset drift, and $\Delta$WER. Fig.~\ref{fig:q2_editing_steps} reveals highly model-dependent, non-uniform trajectory behavior. For F5-TTS, skipping $k=4$ steps retains 99.5\% of the emotion gain while slashing onset drift to 97 ms and improving transcription.

Conversely, for CosyVoice 2, skipping $k=1$ step preserves emotion but fails to reduce drift, whereas skipping $k=3$ steps reduces drift at the expense of emotion strength and WER.
Further early-stop and block-wise ablation analyses also confirm this temporal non-uniformity, which are detailed in Appendix~\ref{app:editing_timing_q2}.

\begin{tcolorbox}[colback=violet!8!white,colframe=black,boxrule=0.7pt,arc=3pt,
  left=6pt,right=6pt,top=5pt,bottom=5pt,boxsep=0pt]
\textbf{Observation 2:} Editability is trajectory- and architecture-dependent: early transport may disrupt timing, while later steps are crucial for effective editing and content recovery.
\end{tcolorbox}

\textbf{Q3: What acoustic attributes are carried by source-to-target velocity transport?}

\begin{wraptable}[11]{r}{0.45\textwidth}
\vspace{-13pt}
\setlength{\intextsep}{0pt}
\setlength{\columnsep}{4pt}
\centering
\begin{threeparttable}
\caption{Cross-speaker emotion editing induces concurrent attribute changes.}
\label{tab:q3_attributes}
\scriptsize
\renewcommand{\arraystretch}{0.9}
\begin{tabular}{@{}lrr@{}}
\toprule
Measure & F5-TTS & CosyVoice 2 \\
\midrule
S-SIM source $\downarrow$ & $-0.403$ & $-0.378$ \\
S-SIM target $\uparrow$ & $0.331$ & $0.342$ \\
Emotion gain (pp) $\uparrow$ & $45.14$ & $52.11$ \\
F0 median gain (semitones) $\uparrow$ & $4.565$ & $7.075$ \\
F0 range gain (semitones) $\uparrow$ & $1.433$ & $2.769$ \\
Joint (/80)\tnote{*} & 70 & 73 \\
\bottomrule
\end{tabular}
\begin{tablenotes}[flushleft]
\item[*] Joint denotes the number of cases where target-speaker similarity increases, source-speaker similarity decreases, and target-emotion probability increases.
\end{tablenotes}
\end{threeparttable}
\end{wraptable}

Building on Q1, we examine if the editing signal isolates emotion when source and target speakers differ.
We edit 80 Neutral utterances (from 20 ESD speakers, fixed text) toward different same-language speakers across four target emotions (Happy, Angry, Sad, Surprise).
We track changes in speaker similarity (S-SIM), target-emotion probability gain (pp), and F0 median/range shifts (semitones).

As Table~\ref{tab:q3_attributes} shows, velocity transport is not attribute-factorized.
Speaker identity and emotion shift jointly (mean emotion gains of 45.14 pp for F5-TTS and 52.11 pp for CosyVoice 2), while F0 median and range also move toward the target reference.
This confirms that cross-speaker transport inevitably entangles speaker and pitch characteristics with emotion. 
Detailed experimental settings and results are provided in Appendix~\ref{app:q3_attributes}.

\begin{tcolorbox}[colback=violet!8!white,colframe=black,boxrule=0.7pt,arc=3pt,
  left=6pt,right=6pt,top=5pt,bottom=5pt,boxsep=0pt]
\textbf{Observation 3:} Velocity transport is not emotion-specific, but carries speaker identity and other attributes, such as pitch characteristics, alongside emotion.
\end{tcolorbox}

\section{SEmoEdit}
\label{sec:semoedit}

\begin{figure}[t]
    \centering
    \includegraphics[width=\linewidth]{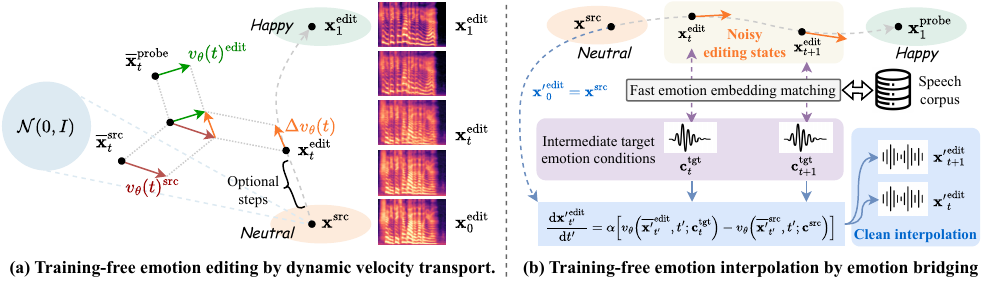}
    \caption{Framework of SEmoEdit for dynamic speech emotion editing.}
    \label{fig:semoedit_framework}
\end{figure}

The observations in Sec.~\ref{sec:analysis} directly inform the design of SEmoEdit.
% Observation~1 shows that conditional velocity differences provide a usable training-free editing signal.
% Observation~2 reveals that this signal is trajectory-dependent, motivating optional skipping of disruptive early steps.
% Observation~3 shows that cross-speaker velocity transport also carries speaker-related attributes, motivating timbre alignment between the source and target emotion references.
Based on these findings, SEmoEdit performs state-dependent velocity transport between speaker-aligned source and target emotion conditions while allowing the transport interval to adapt to the TTS backbone.

\subsection{Dynamic Velocity Transport}
\label{subsec:dynamic_velocity_transport}

\textbf{Coupled emotion queries.}
Given a source acoustic representation
\(\rvx^{\mathrm{src}}\) and source emotion condition
\(\rvc^{\mathrm{src}}\), we construct an editing trajectory
\(\{\rvx_t^{\mathrm{edit}}\}_{t\in[0,1]}\) directly in the acoustic space.
Here, \(t\) follows the CFM convention in Eq.~\ref{eq:cfm_path}, from the noise endpoint to the data endpoint, while
\(\rvx_0^{\mathrm{edit}}=\rvx^{\mathrm{src}}\) serves as the boundary condition of the editing dynamics rather than a sample from \(p_0 = \mathcal{N}(\mathbf{0},\mathbf{I})\).
At each time \(t\), we sample
\(\boldsymbol{\epsilon}_t\sim p_0\) with the same shape as
\(\rvx^{\mathrm{src}}\) and construct a noisy source query
\begin{equation}
    \overline{\rvx}_t^{\mathrm{src}}
    =(1-t)\boldsymbol{\epsilon}_t+t\rvx^{\mathrm{src}}.
    \label{eq:semoedit_source_query}
\end{equation}
We then add the same noise perturbation used for the source query to the current editing state, yielding the coupled target-side query
\begin{equation}
    \overline{\rvx}_t^{\mathrm{edit}}
    =
    \rvx_t^{\mathrm{edit}}
    +
    \left(
    \overline{\rvx}_t^{\mathrm{src}}
    -
    \rvx^{\mathrm{src}}
    \right).
    \label{eq:semoedit_target_query}
\end{equation}
Thus, the source and target-side queries share the same noise perturbation while being anchored at $\rvx^{\mathrm{src}}$ and $\rvx_t^{\mathrm{edit}}$, respectively.
At initialization, they coincide:
\(\overline{\rvx}_0^{\mathrm{edit}}
=\overline{\rvx}_0^{\mathrm{src}}
=\boldsymbol{\epsilon}_0\).

Following Observation~1, we define the instantaneous editing signal as the velocity difference
\begin{equation}
    \Delta\vv_\theta(t)
    =
    \vv_\theta\!\left(
        \overline{\rvx}_t^{\mathrm{edit}},t;
        \rvc^{\mathrm{tgt}}
    \right)
    -
    \vv_\theta\!\left(
        \overline{\rvx}_t^{\mathrm{src}},t;
        \rvc^{\mathrm{src}}
    \right).
    \label{eq:semoedit_velocity_difference}
\end{equation}
Unlike a fixed activation-steering direction \citep{xie2025emosteer,wang2026cocoemo},
\(\Delta\vv_\theta(t)\) is recomputed from the evolving editing state at every step, making the transport more robust and stable.

\textbf{Speaker-aligned emotion conditions.}
Observation~3 shows that cross-speaker velocity transport also carries speaker identity. We therefore align the target-reference timbre with the source speaker before computing Eq.~\ref{eq:semoedit_velocity_difference}:
$\rvc^{\mathrm{tgt}}
=
\mathcal{E}_{\mathrm{vc}}
\left(
\rvc^{\mathrm{tgt}'};
\rvc^{\mathrm{src}}
\right)$,
where \(\mathcal{E}_{\mathrm{vc}}\) converts the target reference to the source-speaker timbre while preserving its emotion, making \(\rvc^{\mathrm{src}}\) and \(\rvc^{\mathrm{tgt}}\) mainly differ in emotion, reducing speaker-dependent components in the velocity difference.

\textbf{Trajectory-aware transport.}
Observation~2 shows that early transport can disrupt temporal structure.
We therefore introduce an optional transport start time
\(\tau\in[0,1]\) and define $g_\tau(t)=\mathbb{I}[t\geq\tau]$,
where \(\tau=0\) recovers full-step transport and
\(\tau>0\) skips the early part of the trajectory.
SEmoEdit then evolves the source utterance according to
\begin{equation}
    \frac{\mathrm{d}\rvx_t^{\mathrm{edit}}}{\mathrm{d}t}
    =
    \alpha\,g_\tau(t)\,
    \mathbb{E}_{\boldsymbol{\epsilon}_t}
    \left[
        \Delta\vv_\theta(t)
        \mid
        \rvx^{\mathrm{src}}
    \right],
    \qquad
    \rvx_0^{\mathrm{edit}}=\rvx^{\mathrm{src}},
    \qquad
    \rvx^{(\alpha)}=\rvx_1^{\mathrm{edit}}.
    \label{eq:dynamic_emotion_transport}
\end{equation}
The strength \(\alpha\in\mathbb{R}\) controls the signed transport magnitude:
\(\alpha=0\) recovers the source utterance,
\(0<\alpha<1\) interpolates toward the target emotion,
\(\alpha=1\) performs the full source-to-target transport.

In practice, the expectation in
Eq.~\ref{eq:dynamic_emotion_transport}
is approximated using \(n_{\mathrm{avg}}\) coupled noise samples.
For editing steps
\(0=t_0<\cdots<t_N=1\), the update becomes
\begin{equation}
    \rvx_{t_{i+1}}^{\mathrm{edit}}
    =
    \rvx_{t_i}^{\mathrm{edit}}
    +
    \alpha\,
    g_\tau(t_i)
    (t_{i+1}-t_i)
    \frac{1}{n_{\mathrm{avg}}}
    \sum_{j=1}^{n_{\mathrm{avg}}}
    \Delta\vv_\theta^{(j)}(t_i).
    \label{eq:dynamic_emotion_transport_discrete}
\end{equation}
Thus, SEmoEdit requires only forward evaluations of the frozen TTS velocity field, with no parameter updates or task-specific optimization.
Fig.~\ref{fig:semoedit_framework}(a) illustrates the evolving editing states.
Appendix~\ref{app:semoedit_algorithm} provides the complete pseudocode.

\subsection{Enabling Stable Interpolation.}

Although the trajectory in Eq.~\ref{eq:dynamic_emotion_transport_discrete} provides intermediate editing states, directly decoding these states does not necessarily produce clean speech.
An intermediate state simultaneously contains source and target acoustic patterns, which may interfere with each other and manifest as audible noise or other artifacts.
In other words, the editing state can represent a meaningful intermediate emotion without itself being a well-formed speech sample.

We address this issue with \textbf{emotion bridging}.
As shown in Fig.~\ref{fig:semoedit_framework}(b), for an intermediate editing state $\rvx_{t_i}^{\mathrm{edit}}$, we first obtain a clean target reference $\rvc_{t}^{\mathrm{tgt}}$ whose emotional representation (emotion2vec embedding) matches that state and align its timbre with the source speaker.
$\rvc_{t}^{\mathrm{tgt}}$ serves as an emotion bridge: rather than using the noisy intermediate state as the output, we apply SEmoEdit again from the original source $\rvx^{\mathrm{src}}$ toward the bridged condition, thereby synthesizing a clean utterance $\mathbf{x'}_t^\mathrm{edit}$ at the corresponding emotion intensity.
The coefficient $\alpha$ controls the emotion strength.

\section{Experiments}

\subsection{SEmoEditBench}
\label{sec:benchmark}

Existing speech editing benchmarks \citep{zhang2026speecheditbench,yan2025ming,ma2026mmae} mainly pair each source utterance with a single instruction, whereas our method requires source--target emotion condition pairs whose velocity difference defines the editing direction. We therefore construct \textbf{SEmoEditBench}, a 600-case paired benchmark with emotion labels and editing instructions for evaluating emotion transfer and non-target attribute preservation.

\textbf{Tasks.}
SEmoEditBench covers \emph{emotion replacement}, \emph{emotion erasure}, and \emph{intensity control}, each under same-dataset same-speaker, same-dataset cross-speaker, and cross-dataset cross-speaker settings.
It contains 320 replacement, 152 erasure, and 128 intensity cases from ESD, IEMOCAP, RAVDESS, and CREMA-D \citep{ZHOU20221,busso2008iemocap,livingstone2018ryerson,CREMA-D2014}.
Details are in Appendix~\ref{app:benchmark_metrics}.

\textbf{Metrics.}
We evaluate our method across four dimensions (details in Appendix~\ref{app:benchmark_metrics}): 
\textbf{1) Effectiveness:} Target emotion probability (TEP) and source emotion suppression (SES) for replacement, neutral probability (NP) and SES for erasure, plus emotion2vec similarity (E-SIM) and directional editing score (DES) when paired targets exist. 
\textbf{2) Intensity Control:} On a 90-case subset, we evaluate EIC-Emb, our newly proposed metric measuring the monotonic movement of emotion2vec embeddings toward targets across five strengths $\alpha\in\{0, 0.25,0.5,0.75,1\}$. 
\textbf{3) Preservation \& Quality:} Assessed via $\Delta$WER ($\Delta$CER for Chinese), S-SIM, and UTMOS \citep{Radford2023whisper,baba2024utmosv2}. 
\textbf{4) Subjective Evaluation:} Speaker (SS-MOS), emotion (ES-MOS), and naturalness (N-MOS) evaluated on 20 sampled cases. All metrics are averaged across cases.

\textbf{Models and Hyperparameters.}
We compare SEmoEdit (applied to three backbones, i.e., F5-TTS, CosyVoice 2, and IndexTTS 2 \citep{zhou2026indextts2}) with representative training-based \citep{yan2025stepaudioeditx,wang2026dotsedit,ma2026auk} and activation-steering methods \citep{xie2025emosteer,wang2026cocoemo}.
For emotion replacement and erasure, we set $\alpha=1$, while for intensity control we use $\alpha\in\{0, 0.25,0.5,0.75,1\}$.
We set $n_{\mathrm{avg}}=1$ and $\tau=0$ for all the main experiments, applying transport over the full generative trajectory.
IndexTTS 2 is used to convert target references to the source speaker's timbre for SEmoEdit.
A speech corpus with 10K samples is created for emotion bridging, and their emotion embeddings are pre-computed for fast matching.

\subsection{Main Results}
\label{sec:main_results}

\begin{table*}[t]
    \centering
    \caption{Main results on the same-dataset same-speaker setting of SEmoEditBench. The top three results for each metric are highlighted in \textbf{bold} with dark, medium, and light pink backgrounds representing the \protect\colorbox{pink!60}{first}, \protect\colorbox{pink!40}{second}, and \protect\colorbox{pink!20}{third} best performances, respectively.}
    \label{tab:main_results}
    % 定义前三名的高亮颜色和加粗宏，需要 \usepackage[table]{xcolor}
    \newcommand{\cfa}[1]{\cellcolor{pink!60}\textbf{#1}} % 第一名：深粉色
    \newcommand{\cfb}[1]{\cellcolor{pink!40}\textbf{#1}} % 第二名：中粉色
    \newcommand{\cfc}[1]{\cellcolor{pink!20}\textbf{#1}} % 第三名：浅粉色
    \renewcommand{\arraystretch}{0.8}
    
    \begin{threeparttable}
        \scriptsize
        \setlength{\tabcolsep}{5.3pt}
        \begin{tabular}{lllccccccc}
            \toprule
            Category & Method & Backbone & \multicolumn{7}{c}{Metrics} \\
            \midrule
            \multicolumn{3}{l}{\textit{Emotion replacement}} & TEP $\uparrow$ & SES $\uparrow$ & E-SIM $\uparrow$ & DES $\uparrow$ & $\Delta$WER $\downarrow$ & S-SIM $\uparrow$ & UTMOS $\uparrow$ \\
            \midrule
            \multirow{3}{*}{\shortstack{Training-\\based}}
                & Step-Audio-EditX & -- & 0.222 & 0.445 & 0.521 & 0.550 & \cfb{-0.048} & 0.567 & \cfa{3.091} \\
                & dots.tts.edit & -- & 0.434 & 0.600 & 0.669 & 0.713 & -0.035 & 0.272 & 2.381 \\
                & Auk & -- & 0.259 & 0.372 & 0.539 & 0.541 & 0.073 & 0.631 & 2.639 \\
            \midrule
            \multirow{5}{*}{\shortstack{Activation\\Steering}}
                & \multirow{2}{*}{CoCoEmo}      & CosyVoice~2 & 0.082 & 0.192 & 0.446 & 0.399 & -0.015 & \cfc{0.715} & \cfb{3.061} \\
                &                                & IndexTTS2  & 0.035 & 0.127 & 0.402 & 0.269 & -0.024 & \cfb{0.776} & 2.659 \\
            \cmidrule(lr){2-10}
                & \multirow{3}{*}{EmoSteer-TTS} & F5-TTS     & 0.081 & 0.345 & 0.448 & 0.477 & 0.608 & 0.623 & 2.545 \\
                &                                & CosyVoice~2 & 0.030 & 0.371 & 0.382 & 0.367 & -0.015 & 0.478 & 2.515 \\
                &                                & IndexTTS2  & 0.025 & 0.054 & 0.377 & 0.226 & \cfc{-0.037} & \cfa{0.788} & 2.666 \\
            \midrule
            \multirow{5}{*}{\textbf{SEmoEdit}}
                & \multirow{3}{*}{\textbf{\shortstack{Audio\\condition}}} & F5-TTS     & \cfc{0.498} & \cfc{0.684} & \cfc{0.704} & \cfc{0.753} & -0.026 & 0.372 & 2.177 \\
                &                                      & CosyVoice~2 & \cfb{0.554} & \cfb{0.694} & \cfb{0.776} & \cfb{0.806} & -0.025 & 0.379 & 2.984 \\
                &                                      & IndexTTS2  & \cfa{0.691} & \cfa{0.767} & \cfa{0.902} & \cfa{0.920} & -0.004 & 0.423 & 2.578 \\
            \cmidrule(lr){2-10}
                & \multirow{2}{*}{\textbf{\shortstack{Text\\condition}}\tnote{$\dagger$}} & CosyVoice~2 & 0.274 & 0.484 & 0.566 & 0.567 & \cfa{-0.059} & 0.610 & \cfc{3.050} \\
                &                                             & IndexTTS2  & 0.354 & 0.506 & 0.629 & 0.634 & 0.013 & 0.594 & 2.572 \\
            \midrule
            \multicolumn{3}{l}{\textit{Emotion erasure}} & NP $\uparrow$ & SES $\uparrow$ & E-SIM $\uparrow$ & DES $\uparrow$ & $\Delta$WER $\downarrow$ & S-SIM $\uparrow$ & UTMOS $\uparrow$ \\
            \midrule
            \multirow{3}{*}{\shortstack{Training-\\based}}
                & Step-Audio-EditX & -- & 0.235 & 0.388 & 0.566 & 0.655 & \cfb{-0.064} & 0.577 & \cfa{3.220} \\
                & dots.tts.edit    & -- & 0.355 & 0.549 & 0.674 & 0.766 & \cfa{-0.065} & 0.304 & 2.685 \\
                & Auk              & -- & 0.229 & 0.311 & 0.528 & 0.582 & 0.014 & \cfb{0.676} & \cfc{2.859} \\
            \midrule
            \multirow{3}{*}{\shortstack{Activation\\Steering}}
                & \multirow{3}{*}{EmoSteer-TTS} & F5-TTS     & 0.202 & 0.393 & 0.616 & 0.736 & 0.030 & 0.598 & 2.593 \\
                &                                & CosyVoice~2 & 0.078 & 0.174 & 0.505 & 0.596 & -0.012 & 0.548 & 2.490 \\
                &                                & IndexTTS2  & 0.038 & 0.070 & 0.377 & 0.298 & -0.041 & \cfa{0.808} & 2.836 \\
            \midrule
            \multirow{5}{*}{\textbf{SEmoEdit}}
                & \multirow{3}{*}{\textbf{\shortstack{Audio\\condition}}} & F5-TTS     & \cfc{0.573} & \cfc{0.700} & \cfc{0.821} & \cfc{0.873} & \cfc{-0.058} & 0.359 & 2.597 \\
                &                                      & CosyVoice~2 & \cfa{0.704} & \cfa{0.754} & \cfb{0.907} & \cfa{0.938} & -0.050 & 0.399 & \cfb{3.144} \\
                &                                      & IndexTTS2  & \cfb{0.682} & \cfb{0.741} & \cfa{0.910} & \cfb{0.934} & -0.030 & 0.424 & 2.774 \\
            \cmidrule(lr){2-10}
                & \multirow{2}{*}{\textbf{\shortstack{Text\\condition}}\tnote{$\dagger$}} & CosyVoice~2 & 0.216 & 0.432 & 0.582 & 0.685 & -0.052 & \cfc{0.607} & \cfc{3.069} \\
                &                                             & IndexTTS2  & 0.104 & 0.309 & 0.471 & 0.504 & -0.050 & \cfb{0.687} & 2.698 \\
            \midrule
            \multicolumn{3}{l}{\textit{Emotion intensity control}} & \multicolumn{2}{c}{EIC-Emb $\uparrow$} & \multicolumn{2}{c}{$\Delta$WER$_{\alpha=1}$ $\downarrow$} & S-SIM$_{\alpha=1}$ $\uparrow$ & \multicolumn{2}{c}{UTMOS$_{\alpha=1}$ $\uparrow$} \\
            \midrule
            \multirow{5}{*}{\shortstack{Activation\\Steering}}
                & \multirow{2}{*}{CoCoEmo}      & CosyVoice~2 & \multicolumn{2}{c}{0.041} & \multicolumn{2}{c}{0.044} & \cfc{0.621} & \multicolumn{2}{c}{\cfa{3.234}} \\
                &                                & IndexTTS2  & \multicolumn{2}{c}{0.017} & \multicolumn{2}{c}{0.333} & \cfa{0.703} & \multicolumn{2}{c}{2.294} \\
            \cmidrule(lr){2-10}
                & \multirow{3}{*}{EmoSteer-TTS} & F5-TTS     & \multicolumn{2}{c}{0.013} & \multicolumn{2}{c}{0.139} & 0.465 & \multicolumn{2}{c}{2.309} \\
                &                                & CosyVoice~2 & \multicolumn{2}{c}{-0.009} & \multicolumn{2}{c}{0.133} & 0.362 & \multicolumn{2}{c}{\cfc{2.575}} \\
                &                                & IndexTTS2  & \multicolumn{2}{c}{-0.005} & \multicolumn{2}{c}{0.400} & \cfb{0.683} & \multicolumn{2}{c}{2.355} \\
            \midrule
            \multirow{5}{*}{\textbf{SEmoEdit}}
                & \multirow{3}{*}{\textbf{\shortstack{Audio\\condition}}} & F5-TTS     & \multicolumn{2}{c}{\cfb{0.175}} & \multicolumn{2}{c}{\cfc{0.039}} & 0.307 & \multicolumn{2}{c}{1.967} \\
                &                                      & CosyVoice~2 & \multicolumn{2}{c}{\cfc{0.172}} & \multicolumn{2}{c}{\cfb{0.017}} & 0.362 & \multicolumn{2}{c}{2.499} \\
                &                                      & IndexTTS2  & \multicolumn{2}{c}{\cfa{0.192}} & \multicolumn{2}{c}{0.411} & 0.336 & \multicolumn{2}{c}{2.355} \\
            \cmidrule(lr){2-10}
                & \multirow{2}{*}{\textbf{\shortstack{Text\\condition}}\tnote{$\dagger$}} & CosyVoice~2 & \multicolumn{2}{c}{0.144} & \multicolumn{2}{c}{\cfa{-0.017}} & 0.553 & \multicolumn{2}{c}{\cfb{3.080}} \\
                &                                             & IndexTTS2  & \multicolumn{2}{c}{0.108} & \multicolumn{2}{c}{0.394} & 0.472 & \multicolumn{2}{c}{2.458} \\
            \bottomrule
        \end{tabular}
        \begin{tablenotes}
            \scriptsize
            \item[$\dagger$] The text condition is an emotion instruction, e.g., \emph{Speak with a happy tone.}.
        \end{tablenotes}
    \end{threeparttable}
\end{table*}

\textbf{Comparison with training-based methods.}
As shown in Table~\ref{tab:main_results}, audio-conditioned SEmoEdit consistently outperforms all training-based baselines in editing success without task-specific training. Even its weakest backbone surpasses the strongest baseline on every emotion-related metric for replacement (e.g., TEP 0.498 vs. 0.434; DES 0.753 vs. 0.713) and erasure (NP 0.573 vs. 0.355; DES 0.873 vs. 0.766), while the best variants reach 0.691/0.920 TEP/DES for replacement and 0.704/0.938 NP/DES for erasure. These gains retain near-zero or negative \(\Delta\)WER and additionally support continuous intensity control. Overall, pretrained generative dynamics provide stronger emotion-editing capability than dedicated editors trained on large-scale task-specific data.
% We encourage the readers to listen to the audio samples in our demo page\footnote{\url{https://semoedit.github.io}} for a more comprehensive evaluation of the editing quality.

\textbf{Comparison with activation steering methods.}
Audio-conditioned SEmoEdit consistently outperforms fixed activation steering across all three backbones.
For replacement and erasure, every SEmoEdit variant surpasses all steering baselines on all four emotion-related metrics.
The advantage is also clear for intensity control, where EIC-Emb reaches 0.175, 0.172, and 0.192 on F5-TTS, CosyVoice~2, and IndexTTS2, compared with 0.013, 0.041, and 0.017 for their strongest steering counterparts.
While SEmoEdit yields lower S-SIM, these consistent gains demonstrate that state-dependent velocity transport provides a more effective editing signal than fixed activation directions.

\textbf{Human Evaluation.}
Table~\ref{tab:subjective_results} (15 human evaluators) confirms SEmoEdit's stronger perceived editing capability. SEmoEdit achieves the best emotion-replacement score (ES-MOS 4.22 with IndexTTS2), the top three erasure scores, and higher intensity-control scores (EIC-MOS 2.80--3.65) than CoCoEmo (1.65) and EmoSteer-TTS (1.30). Although the baselines sometimes obtain higher SS-MOS or N-MOS, their low editing scores suggest little emotion changes. SEmoEdit therefore provides a better trade-off between edit effectiveness and output quality.

\begin{table*}[t]
    \centering
    \caption{Subjective evaluation results. The \protect\colorbox{pink!60}{first}, \protect\colorbox{pink!40}{second}, and \protect\colorbox{pink!20}{third} place are highlighted.}
    \label{tab:subjective_results}
    \newcommand{\sfa}[1]{\cellcolor{pink!60}\textbf{#1}}
    \newcommand{\sfb}[1]{\cellcolor{pink!40}\textbf{#1}}
    \newcommand{\sfc}[1]{\cellcolor{pink!20}\textbf{#1}}
    \renewcommand{\arraystretch}{0.8}
    \begin{threeparttable}
        \scriptsize
        \setlength{\tabcolsep}{2.7pt}
        \resizebox{\textwidth}{!}{%
        \begin{tabular}{lllccccccc}
            \toprule
            & & & \multicolumn{3}{c}{Emotion replacement} & \multicolumn{3}{c}{Emotion erasure} & Intensity control \\
            \cmidrule(lr){4-6} \cmidrule(lr){7-9} \cmidrule(lr){10-10}
            Category & Method & Backbone & SS-MOS $\uparrow$ & ES-MOS $\uparrow$ & N-MOS $\uparrow$ & SS-MOS $\uparrow$ & ES-MOS $\uparrow$ & N-MOS $\uparrow$ & EIC-MOS $\uparrow$ \\
            \midrule
            \multirow{3}{*}{\shortstack{Training-\\based}}
                & Step-Audio-EditX & -- & \sfc{3.67} & 2.88 & 3.75 & \sfb{4.20} & 2.35 & 4.00 & -- \\
                & Auk              & -- & 3.38 & 2.80 & 3.52 & \sfc{4.00} & 2.95 & \sfc{4.05} & -- \\
                & dots.tts.edit    & -- & 3.52 & \sfb{3.27} & 3.77 & 3.55 & 3.15 & 3.65 & -- \\
            \midrule
            \multirow{2}{*}{\shortstack{Activation\\Steering}}
                & CoCoEmo      & IndexTTS2 & \sfa{4.42} & 1.88 & \sfb{3.98} & -- & -- & -- & 1.65 \\
            \cmidrule(lr){2-10}
                & EmoSteer-TTS & IndexTTS2 & \sfb{4.35} & 1.75 & \sfa{4.12} & \sfa{4.50} & 1.80 & \sfa{4.25} & 1.30 \\
            \midrule
            \multirow{3}{*}{SEmoEdit}
                & \multirow{3}{*}{Ours} & F5-TTS     & 2.88 & \sfc{3.23} & 3.25 & 2.85 & \sfc{3.65} & 3.80 & \sfc{2.80} \\
                &                       & CosyVoice~2 & 3.02 & 3.20 & 3.42 & 3.45 & \sfb{4.30} & \sfb{4.15} & \sfb{3.25} \\
                &                       & IndexTTS2  & 3.45 & \sfa{4.22} & \sfc{3.88} & 3.55 & \sfa{4.40} & \sfb{4.15} & \sfa{3.65} \\
            \bottomrule
        \end{tabular}
        }
    \end{threeparttable}
\end{table*}

\begin{figure*}[t]
    \centering
    \begin{minipage}[t]{0.525\textwidth}
        \centering
        \includegraphics[width=\linewidth]{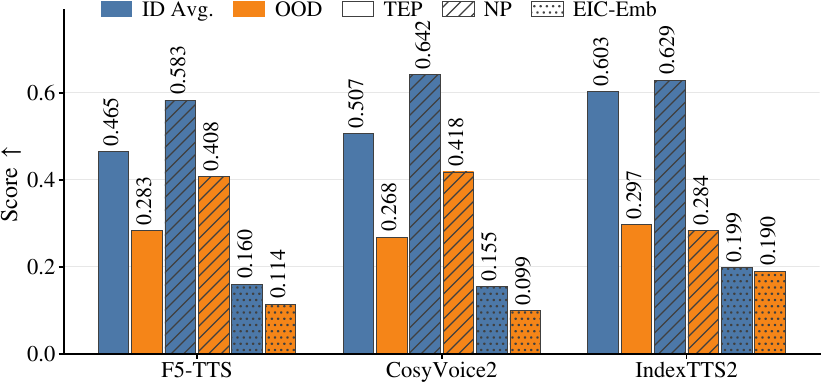}
    \end{minipage}
    \hfill
    \begin{minipage}[t]{0.455\textwidth}
        \centering
        \includegraphics[width=\linewidth]{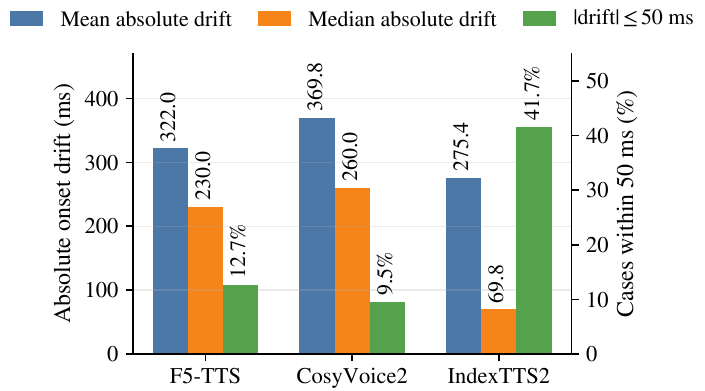}
    \end{minipage}
    \caption{\textbf{Left:} SEmoEdit ID/OOD generalization across backbones. \textbf{Right:} Onset drift over 600 cases; lower mean and median absolute drift are better, while a higher fraction within 50 ms is better.}
    \label{fig:generalization_results}
    \label{fig:onset_drift_600cases}
\end{figure*}

\textbf{Generalization to out-of-distribution cases.}
The left panel of Fig.~\ref{fig:generalization_results} shows that SEmoEdit remains effective in the challenging cross-dataset, cross-speaker setting without retraining, outperforming most ID baselines in Table~\ref{tab:main_results}.
For intensity control, IndexTTS2 retains 95\% of its ID-average score (0.190 vs. 0.199), whereas F5-TTS and CosyVoice~2 decline to some extent.
Replacement and erasure also show TEP/NP drops. Overall, dynamic velocity transport generalizes across corpora and speakers without retraining, but continuous-control robustness depends on the backbone and cross-domain categorical emotion transfer remains a key limitation. Full results are in Appendix~\ref{app:complete_benchmark_results}.

\subsection{Ablation Study}
\label{sec:ablation_study}

\textbf{Analysis of onset drifting.}
We further evaluate onset drift across all three backbones using 600 SEmoEditBench cases. As Fig.~\ref{fig:onset_drift_600cases} shows, IndexTTS2 best preserves timing with a 69.8-ms median drift (41.7\% $\le$50 ms).
Although long tail cases inflates IndexTTS2's mean, CosyVoice~2 yields the highest mean and median, confirming its stronger emotion-temporal coupling.
This contrast reflects architecture designs: F5-TTS establishes timing jointly along fixed text length, with no explicit duration control; CosyVoice~2 relies on unaligned autoregressive semantic tokens, leaving its flow prior to impose a target-specific temporal scaffold; IndexTTS2 instead enforces a duration-controlled frame layout before flow matching, substantially reducing onset drift.

% \begin{tcolorbox}[colback=violet!8!white,colframe=black,boxrule=0.7pt,arc=3pt,
%   left=6pt,right=6pt,top=5pt,bottom=5pt,boxsep=0pt]
% \textbf{Takeaway 1:} Onset-stable editing benefits from explicit duration or frame allocation.
% \end{tcolorbox}

\textbf{Analysis of noise.}
\begin{figure}[t]
    \centering
    \includegraphics[width=\textwidth]{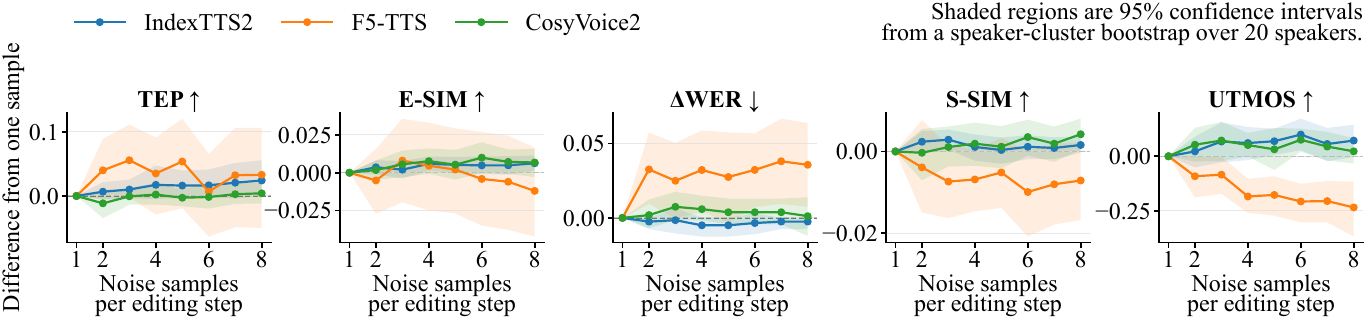}
    \caption{Sensitivity to the number of coupled noise samples per editing step.}
    \label{fig:noise_sampling_sensitivity}
\end{figure}
Fig.~\ref{fig:noise_sampling_sensitivity} shows that increasing \(n_{\mathrm{avg}}\) from 1 to 8 has little effect on CosyVoice~2 and IndexTTS2. F5-TTS is more sensitive, but gains remain inconsistent (at \(n_{\mathrm{avg}}=8\): TEP \(+0.033\), \(\Delta\)WER \(+0.036\), UTMOS \(-0.234\)). Since additional samples provide no consistent benefit while increasing velocity-query cost up to \(8\times\), we use \(n_{\mathrm{avg}}=1\) throughout all experiments.

% \begin{tcolorbox}[colback=violet!8!white,colframe=black,boxrule=0.7pt,arc=3pt,
%   left=6pt,right=6pt,top=5pt,bottom=5pt,boxsep=0pt]
% \textbf{Takeaway 2:} Noise averaging is not necessary for velocity-transport-based emotion editing.
% \end{tcolorbox}

\begin{wrapfigure}[3]{r}{0.31\textwidth}
    \vspace{-\intextsep}
    \centering
    \includegraphics[width=\linewidth]{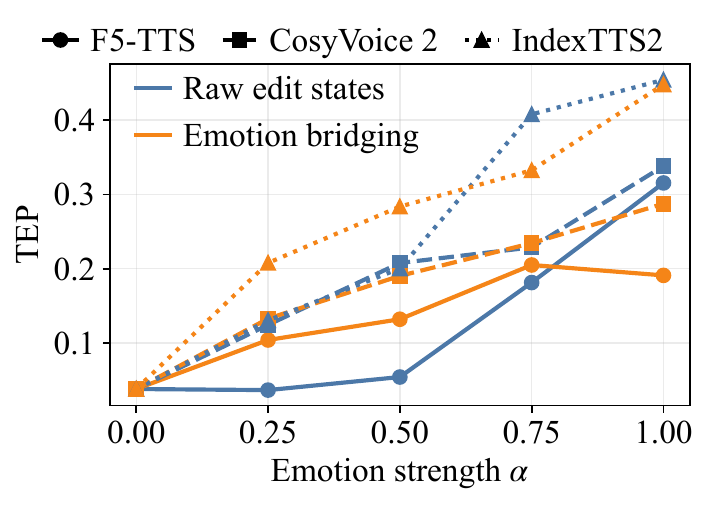}
    \vspace{-15pt}
    \caption{Mean TEP trajectories before and after bridging.}
    \label{fig:emotion_bridging_tep}
    \vspace{-\intextsep}
\end{wrapfigure}

\noindent\begin{minipage}[t]{0.65\textwidth}
    \vspace{-12pt}
    % \vspace{0pt}
    \centering
    \captionof{table}{Objective metrics before and after emotion bridging, averaged over 90 intensity editing cases.}
    \resizebox{\linewidth}{!}{%
\renewcommand{\arraystretch}{0.8}
\scriptsize
\begin{tabular}{@{}llccc@{}}
\toprule
Backbone & Variant & WER/CER $\downarrow$ & S-SIM $\uparrow$ & UTMOS $\uparrow$ \\
\midrule
\multirow{2}{*}{F5-TTS} & Raw & 0.342 & 0.418 & 2.151 \\
 & Bridged & \textbf{0.205} & \textbf{0.431} & \textbf{2.411} \\
\addlinespace[2pt]
\multirow{2}{*}{CosyVoice~2} & Raw & 0.360 & 0.349 & 2.134 \\
 & Bridged & \textbf{0.201} & \textbf{0.457} & \textbf{2.780} \\
\addlinespace[2pt]
\multirow{2}{*}{IndexTTS2} & Raw & 0.823 & 0.379 & 1.918 \\
 & Bridged & \textbf{0.603} & \textbf{0.440} & \textbf{2.652} \\
\bottomrule
\end{tabular}%
}

    \label{tab:emotion_bridging_metrics}
\end{minipage}
% \par

\textbf{Analysis of the emotion bridging.}
Table~\ref{tab:emotion_bridging_metrics} shows emotion bridging consistently improves acoustic stability across all backbones. It reduces WER/CER, increases UTMOS, and improves S-SIM, notably for CosyVoice~2. This suggests the second pass removes intermediate acoustic artifacts, yielding clearer, speaker-consistent speech.
As Fig.~\ref{fig:emotion_bridging_tep} shows, bridging maintains highly stable intensity separation across most levels, exhibiting only minor fluctuations at high strengths. Thus, it effectively acts as a fidelity regularizer while largely preserving the interpolation range, intelligibility, timbre, and naturalness.

\textbf{Audio condition vs. text condition.}
As shown in Table~\ref{tab:main_results}, replacing audio conditions with text instructions sharply reduces replacement TEP from 0.554/0.691 to 0.274/0.354 on CosyVoice~2/IndexTTS2, with similar degradation in erasure and intensity control. This indicates that current instruction-conditioned TTS models cannot reliably induce the target-emotion velocity difference required by SEmoEdit, despite better speaker preservation.

% \textbf{Can we use fewer steps for transport?}

\subsection{Takeaways for Real Applications and Future Directions}

Based on our experiments and findings, we want to highlight several key insights and potential future directions:
(1) Flow-based editing techniques can be applied to acoustic flows, but the temporal nature of audio requires careful consideration.
(2) The choice of backbone TTS model significantly affects the editing performance, with models that have explicit duration control (e.g., IndexTTS2) showing better stability.
(3) Noise averaging does not provide consistent benefits.
(4) Emotion bridging indicates that multi-pass editing strategies may be beneficial for complex tasks.
(5) Text-conditioned training-free editing currently lags behind, highlighting the need for improved instruction-conditioned TTS models that can better capture the desired emotional transformations.

\section{Related Work}
\label{sec:related_work}

\textbf{Emotion-Conditioned Speech Synthesis} generates new speech from prompts or labels. These methods offer prompt-driven control (PromptTTS \citep{PromptTTS}, ControlSpeech \citep{ji2025controlspeech}, EmoVoice \citep{yang2025emovoice}), continuous intensity tuning (EmoSphere++ \citep{Cho2025EmoSphere++}), word-level alignment (WeSCon \citep{wang2025word}), or timbre disentanglement (IndexTTS2 \citep{zhou2026indextts2}). However, they cannot edit existing audio.
\textbf{Training-Based Emotion Editing} modifies source speech via mask-based inpainting (Emo-CampNet \citep{wang2024emocampnet}), instruction-driven LMs (Step-Audio-EditX \citep{yan2025stepaudioeditx}, SpeechEdit \citep{pei2026speechedit}), caption rewriting (Bagpiper-Edit \citep{gong2026bagpiper}), transcript-grounded span editing (dots.tts.edit \citep{wang2026dotsedit}), or voice-level tuning (VoiceDesigner \citep{hai2026voicedesigner}). Despite their flexibility, these rely on expensive annotations, dedicated training, or post-training.
\textbf{Inference-Time Representation Steering} manipulates representations without retraining via activation steering (EmoSteer-TTS \citep{xie2025emosteer}, CoCoEmo \citep{wang2026cocoemo}), lightweight interventions (EmoShift \citep{zhou2026emoshift}), or modifying SAE features \citep{du2026sparse}. Unlike these methods that primarily regenerate speech or steer via fixed offsets, SEmoEdit dynamically modulates velocity transport to edit existing audio, unifying emotion replacement, erasure, and interpolation.
% Table~\ref{tab:method_capability_comparison} compares the capabilities of our SEmoEdit approach with representative editing methods.

\section{Conclusion}
\label{sec:conclusion}

This work presents the first systematic study of training-free emotion editability in pre-trained flow-matching and hybrid TTS models. Driven by our in-depth analysis and three key observations that reveal the potential and limitations of repurposing speech flows, we introduce the SEmoEdit framework. Extensive comparison and evaluations on SEmoEditBench demonstrate that SEmoEdit serves as a highly practical alternative to training-based approaches, achieving competitive, and often superior, performance without requiring parameter updates. We hope these insights inspire broader exploration of inference-time speech editing within foundational audio models.

\bibliographystyle{arxiv}
\bibliography{arxiv}

\appendix

\section{SEmoEdit Algorithm}
\label{app:semoedit_algorithm}

Algorithm~\ref{alg:semoedit} summarizes the discrete dynamic velocity transport in Eq.~\ref{eq:dynamic_emotion_transport_discrete}.
The target reference is first converted to the source-speaker timbre; when the two references are already speaker-aligned, this conversion is the identity operation.
For backbones that prepend reference frames, each velocity evaluation retains its branch-specific prefix, and the difference below is computed only over the shared generated region.

\begin{algorithm}[H]
\caption{SEmoEdit via dynamic velocity transport}
\label{alg:semoedit}
\begin{algorithmic}[1]
\Require Frozen velocity field $\vv_\theta$; source representation $\rvx^{\mathrm{src}}$; source and target-reference conditions $\rvc^{\mathrm{src}},\rvc^{\mathrm{tgt}'}$; speaker alignment operator $\mathcal{E}_{\mathrm{vc}}$; time grid $0=t_0<\cdots<t_N=1$; strength $\alpha$; start time $\tau$; noise samples per step $n_{\mathrm{avg}}$
\Ensure Edited representation $\rvx^{(\alpha)}$
\State $\rvc^{\mathrm{tgt}} \gets \mathcal{E}_{\mathrm{vc}}(\rvc^{\mathrm{tgt}'};\rvc^{\mathrm{src}})$
\State $\rvx_{t_0}^{\mathrm{edit}} \gets \rvx^{\mathrm{src}}$
\For{$i=0,\ldots,N-1$}
    \State $\Delta\vv_i \gets \mathbf{0}$
    \For{$j=1,\ldots,n_{\mathrm{avg}}$}
        \State Sample $\boldsymbol{\epsilon}_{i,j}\sim p_0$ with the same shape as $\rvx^{\mathrm{src}}$
        \State $\overline{\rvx}_{i,j}^{\mathrm{src}} \gets (1-t_i)\boldsymbol{\epsilon}_{i,j}+t_i\rvx^{\mathrm{src}}$
        \State $\overline{\rvx}_{i,j}^{\mathrm{edit}} \gets \rvx_{t_i}^{\mathrm{edit}}+\overline{\rvx}_{i,j}^{\mathrm{src}}-\rvx^{\mathrm{src}}$
        \State $\Delta\vv_i \gets \Delta\vv_i
        +\vv_\theta(\overline{\rvx}_{i,j}^{\mathrm{edit}},t_i;\rvc^{\mathrm{tgt}})
        -\vv_\theta(\overline{\rvx}_{i,j}^{\mathrm{src}},t_i;\rvc^{\mathrm{src}})$
    \EndFor
    \State $\rvx_{t_{i+1}}^{\mathrm{edit}} \gets \rvx_{t_i}^{\mathrm{edit}}
    +\alpha\,\mathbb{I}[t_i\geq\tau](t_{i+1}-t_i)\Delta\vv_i/n_{\mathrm{avg}}$
\EndFor
\State \Return $\rvx^{(\alpha)}\gets\rvx_{t_N}^{\mathrm{edit}}$
\end{algorithmic}
\end{algorithm}

\section{Details of the Editability Diagnostic}
\label{app:editability_diagnostic}

\subsection{Experiment Details for Q1}
\label{app:editability_diagnostic_q1}

\textbf{Experimental setup.}
We construct 80 neutral-to-emotion editing cases from the Emotional Speech Database (ESD), covering all 20 speakers (ten Mandarin and ten English) and four target emotions: Happy, Angry, Sad, and Surprise. For each speaker, four distinct parallel text groups are assigned to the four target emotions. Each case contains a neutral reference $R_s$ and a target-emotion reference $R_t$ spoken by the same speaker with the same text, together with a separate synthesis transcript that remains fixed during editing. Both models are evaluated on the same cases, giving 20 cases per emotion and 40 per language. We use frozen F5-TTS v1 Base and CosyVoice~2-0.5B with their supplied pre-trained vocoders; no additional training is performed.

F5-TTS uses 32 Euler integration steps with its supplied time schedule and sway-sampling coefficient $-1$, while CosyVoice 2 uses 10 Euler steps with its supplied cosine schedule and 32-bit floating-point computation. Classifier-free guidance uses scales $\gamma=2$ for F5-TTS and $\gamma=0.7$ for CosyVoice 2. Both editors use strength $\alpha=1$, one noise sample per step, and the full interval $[0,1]$. Case selection uses seed 42, and generation uses seed $42+j$ for zero-indexed case $j$.

\textbf{Editing procedure.}
Following Sec.~\ref{subsec:dynamic_velocity_transport}, $\rvx^{\mathrm{src}}$ is the length-aligned source-generated mel segment, $\rvx_t^{\mathrm{edit}}$ is the evolving edited segment, and $\rvc^{\mathrm{src}},\rvc^{\mathrm{tgt}}$ are constructed from $R_s,R_t$. Each model input contains a reference prefix followed by the generated segment. We initialize $\rvx_{t_0}^{\mathrm{edit}}=\rvx^{\mathrm{src}}$ and use evaluation times $0=t_0<\cdots<t_K=1$, with $K=32$ for F5-TTS and $K=10$ for CosyVoice 2. At each step, the two branches share a fresh noise sample $\boldsymbol{\epsilon}_{t_i}\sim p_0$. The Euler discretization is
\begin{equation}
    \begin{gathered}
        \overline{\rvx}_{t_i}^{\mathrm{src}}=(1-t_i)\boldsymbol{\epsilon}_{t_i}+t_i\rvx^{\mathrm{src}},\qquad
        \overline{\rvx}_{t_i}^{\mathrm{edit}}=\rvx_{t_i}^{\mathrm{edit}}+\overline{\rvx}_{t_i}^{\mathrm{src}}-\rvx^{\mathrm{src}}, \\
        \rvx_{t_{i+1}}^{\mathrm{edit}}=\rvx_{t_i}^{\mathrm{edit}}+\alpha(t_{i+1}-t_i)\left[
        \vv_\theta\!\left(\overline{\rvx}_{t_i}^{\mathrm{edit}},t_i;\rvc^{\mathrm{tgt}}\right)
        -\vv_\theta\!\left(\overline{\rvx}_{t_i}^{\mathrm{src}},t_i;\rvc^{\mathrm{src}}\right)\right].
    \end{gathered}
    \label{eq:editability_probe_discrete}
\end{equation}
Each velocity evaluation retains its branch-specific reference prefix, which is removed before subtracting the equal-length generated-region velocities. Only the generated segment is updated, and the final state $\rvx^{\mathrm{edit}}=\rvx_{t_K}^{\mathrm{edit}}$ is decoded with the model's vocoder.

\begin{figure}[!htbp]
    \centering
    \includegraphics[width=\textwidth]{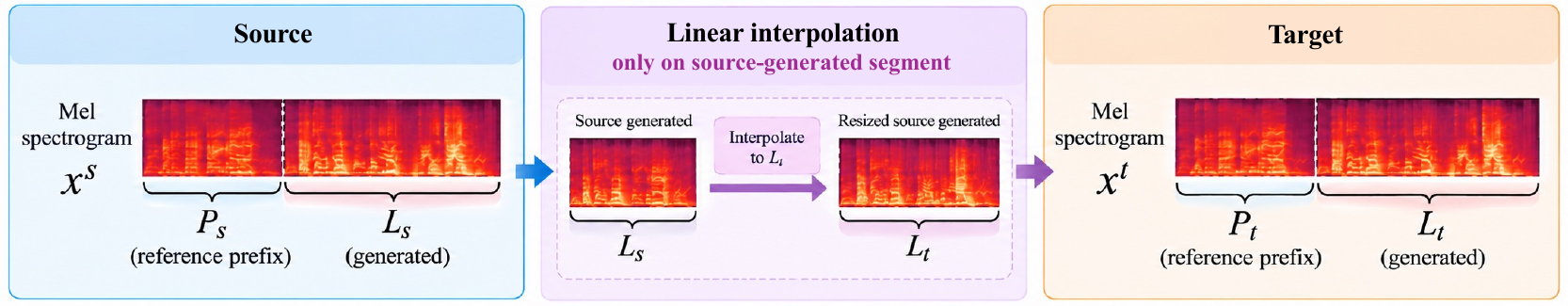}
    \caption{F5-TTS conditioning. The source-generated mel segment is linearly interpolated to the length of the target-generated mel segment.}
    \label{fig:f5ttsconditioning}
\end{figure}

\textbf{F5-TTS conditioning.}
The source and target branches use the mel spectrograms of $R_s$ and $R_t$, respectively, while each branch's text tokens contain its reference transcript followed by the shared synthesis transcript. We synthesize the two branches separately and linearly interpolate only the source-generated mel segment to the target-generated length, as shown in Fig.~\ref{fig:f5ttsconditioning}. Both reference mel conditions and their state prefixes retain their original lengths; the acoustic conditions are zero-padded over the generated region.

\begin{figure}[!htbp]
    \centering
    \includegraphics[width=\textwidth]{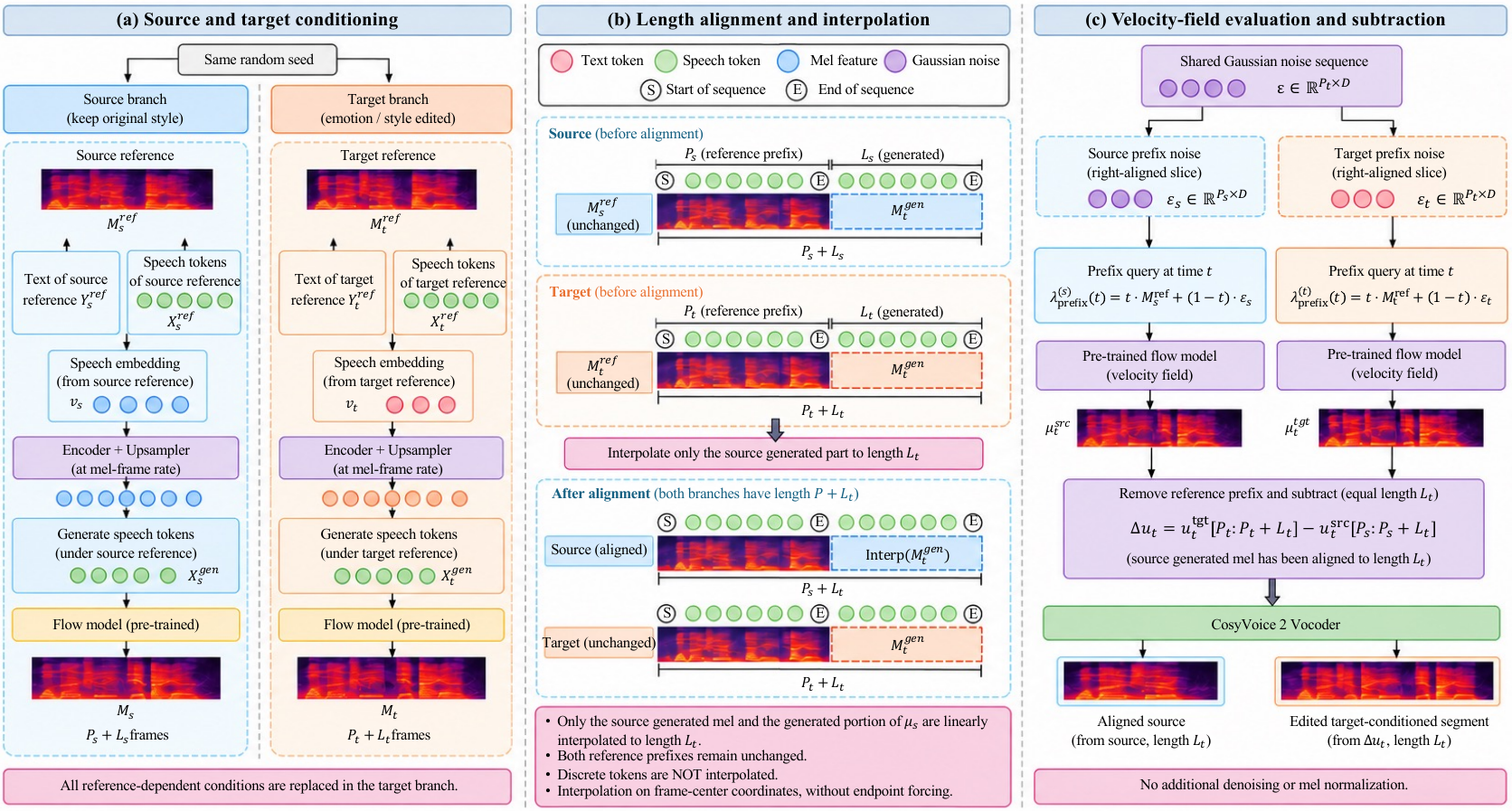}
    \caption{CosyVoice 2 conditioning. Each branch uses its own reference mel and speaker embedding, together with the generated mel segment under that reference. The source branch's generated mel is linearly interpolated to the target length, while the target branch remains unchanged. The two branches share a Gaussian noise sequence of length $\max(P_s,P_t)$, which is right-aligned to each reference prefix.}
    \label{fig:CosyVoice2conditioning}
\end{figure}

\textbf{CosyVoice 2 conditioning.}
As shown in Fig.~\ref{fig:CosyVoice2conditioning}, source and target sequences are synthesized separately using their respective references and the same random seed. Each branch uses the text and discrete speech tokens of its own reference, together with the speech tokens generated under that reference; these are encoded and upsampled to a continuous representation $\mu$ at the mel-frame rate. The flow model additionally receives the corresponding speaker embedding and reference mel, so the target branch replaces all reference-dependent conditions. Let $P_s,P_t$ denote the source and target prefix lengths and $L_s,L_t$ their generated lengths. We interpolate only the source-generated mel and the generated portion of $\mu_s$ to $L_t$, leaving the target representation, both reference prefixes, and all discrete tokens unchanged. Interpolation uses frame-center coordinates without endpoint forcing. The resulting branch lengths are $P_s+L_t$ and $P_t+L_t$; after evaluating the full velocity fields, we remove each reference prefix and subtract the equal-length generated portions. Prefix noises are right-aligned slices of a shared Gaussian sequence of length $\max(P_s,P_t)$. Aligned-source and edited segments are decoded with the same pre-trained vocoder and random state, without additional denoising or mel normalization.

\textbf{Evaluation.}
We evaluate waveforms decoded from $\rvx^{\mathrm{src}}$ and $\rvx^{\mathrm{edit}}$. Let $\rvx^{\mathrm{src,unaligned}}$ denote the source mel before length interpolation; for F5-TTS it equals $\rvx^{\mathrm{src}}$. Target-emotion probability is computed with the pre-trained emotion2vec-plus-large\footnote{\url{https://huggingface.co/emotion2vec/emotion2vec_plus_large}} classifier by applying a softmax over Neutral, Happy, Angry, Sad, and Surprise. WER is computed against the synthesis transcript using Whisper-large-v3\footnote{\url{https://huggingface.co/openai/whisper-large-v3}}. Text is lowercased, stripped of Unicode punctuation, whitespace-normalized, and segmented with Jieba\footnote{\url{https://github.com/fxsjy/jieba}} for Mandarin. All evaluation audio is converted to mono and resampled to 16 kHz without additional loudness normalization. Metrics are averaged over cases.

Probability and WER compare decoded $\rvx^{\mathrm{src}}$ and $\rvx^{\mathrm{edit}}$. Speaker similarity (S-SIM) instead uses $\rvx^{\mathrm{src,unaligned}}$ as the baseline to include the complete pipeline and compares each waveform with an emotion-matched reference:
\begin{equation}
 \begin{gathered}
 S_{\mathrm{before}}=\cos\!\left(E(\mathcal{V}(\rvx^{\mathrm{src,unaligned}})),E(R_s)\right),\\
 S_{\mathrm{after}}=\cos\!\left(E(\mathcal{V}(\rvx^{\mathrm{edit}})),E(R_t)\right),\qquad
 \Delta S=S_{\mathrm{after}}-S_{\mathrm{before}}.
 \end{gathered}
 \label{eq:q1_matched_similarity}
\end{equation}
Here, $E$ is an ECAPA-TDNN\footnote{\url{https://huggingface.co/speechbrain/spkrec-ecapa-voxceleb}} speaker encoder and $\mathcal{V}$ is the model's waveform decoder. Because the speaker encoder can remain sensitive to emotion, emotion-matched references reduce the mismatch caused by comparing emotional speech only against a neutral reference. Accordingly, $\Delta S$ should be interpreted as reference-based speaker-similarity change rather than a direct measure of identity loss.

\begin{table}[!htbp]
\centering
\footnotesize
\caption{Target-emotion probability before and after editing. Values are percentages and $\Delta$ is in percentage points (pp). CosyVoice uses the aligned-source baseline.}
\label{tab:q1_probability_detail}
\scriptsize
\begin{tabular}{llrrr}
\toprule
Model & Emotion & Before (\%, $\uparrow$) & After (\%, $\uparrow$) & $\Delta$ (pp, $\uparrow$) \\
\midrule
F5-TTS & Happy & 5.0456 & 64.8676 & +59.8219 \\
 & Angry & 19.9294 & 77.8099 & +57.8805 \\
 & Sad & 19.8356 & 60.8144 & +40.9788 \\
 & Surprise & 0.2028 & 39.6302 & +39.4274 \\
 & Mean (80 cases) & 11.2534 & 60.7805 & +49.5272 \\
\midrule
CosyVoice 2 & Happy & 10.0841 & 70.5735 & +60.4894 \\
 & Angry & 20.0920 & 86.6830 & +66.5910 \\
 & Sad & 19.6899 & 54.9719 & +35.2820 \\
 & Surprise & $2.4802\!\times\!10^{-5}$ & 61.0794 & +61.0794 \\
 & Mean (80 cases) & 12.4665 & 68.3269 & +55.8604 \\
\midrule
Both models & Mean (160 observations) & 11.8599 & 64.5537 & +52.6938 \\
\bottomrule
\end{tabular}
\end{table}

\begin{table}[!htbp]
    \centering
    \footnotesize
\caption{Per-case-averaged WER before and after editing. Positive changes indicate degradation. CosyVoice uses the aligned-source baseline; its whole-pipeline comparison is reported in the text.}
\label{tab:q1_wer_detail}
\scriptsize
\begin{tabular}{llrrr}
\toprule
Model & Emotion & Before (\%, $\downarrow$) & After (\%, $\downarrow$) & $\Delta$ (pp, $\downarrow$) \\
\midrule
F5-TTS & Happy & 18.0595 & 15.3095 & -2.7500 \\
 & Angry & 20.3591 & 9.4702 & -10.8889 \\
 & Sad & 9.7123 & 8.9385 & -0.7738 \\
 & Surprise & 12.4702 & 16.5575 & +4.0873 \\
 & Mean (80 cases) & 15.1503 & 12.5689 & -2.5813 \\
\midrule
CosyVoice 2 & Happy & 8.8095 & 13.8452 & +5.0357 \\
 & Angry & 20.6627 & 10.2619 & -10.4008 \\
 & Sad & 4.7500 & 9.7202 & +4.9702 \\
 & Surprise & 6.0813 & 8.8075 & +2.7262 \\
 & Mean (80 cases) & 10.0759 & 10.6587 & +0.5828 \\
\midrule
Both models & Mean (160 observations) & 12.6131 & 11.6138 & -0.9993 \\
\bottomrule
\end{tabular}
\end{table}

\begin{table}[t]
    \footnotesize
    \centering
\caption{Emotion-matched S-SIM defined in Eq.~\ref{eq:q1_matched_similarity}. Before uses the neutral reference; after uses the target-emotion reference. CosyVoice before is the unedited generated speech before length alignment, so changes include alignment. $\Delta$ is an absolute cosine difference.}
\label{tab:q1_similarity_detail}
\scriptsize
\begin{tabular}{llrrr}
\toprule
Model & Emotion & Before $\uparrow$ & After $\uparrow$ & $\Delta$ $\uparrow$ \\
\midrule
F5-TTS & Happy & 0.6522 & 0.5966 & -0.0556 \\
 & Angry & 0.6629 & 0.5695 & -0.0934 \\
 & Sad & 0.6591 & 0.6186 & -0.0405 \\
 & Surprise & 0.6878 & 0.5499 & -0.1379 \\
 & Mean (80 cases) & 0.6655 & 0.5836 & -0.0819 \\
\midrule
CosyVoice 2 & Happy & 0.6454 & 0.5716 & -0.0738 \\
 & Angry & 0.6501 & 0.5373 & -0.1127 \\
 & Sad & 0.6362 & 0.6444 & +0.0082 \\
 & Surprise & 0.6519 & 0.5589 & -0.0930 \\
 & Mean (80 cases) & 0.6459 & 0.5781 & -0.0678 \\
\midrule
Both models & Mean (160 observations) & 0.6557 & 0.5808 & -0.0748 \\
\bottomrule
\end{tabular}
\end{table}

\textbf{Results.}
Tables~\ref{tab:q1_probability_detail}, \ref{tab:q1_wer_detail}, and \ref{tab:q1_similarity_detail} provide the complete per-emotion results and model averages. Both models exhibit pronounced target-emotion probability gains: 49.53 pp for F5-TTS and 55.86 pp for CosyVoice 2. WER decreases by 2.58 pp for F5-TTS but increases by 0.58 pp for CosyVoice 2; the per-emotion results show that these averages mask heterogeneous changes across targets. Mean emotion-matched S-SIM decreases by 0.0819 for F5-TTS and 0.0678 for CosyVoice 2. Percentile 95\% confidence intervals for these S-SIM differences are $[-0.1013,-0.0605]$ and $[-0.0833,-0.0520]$, obtained from 10,000 bootstrap samples of the 20 speaker clusters with seed 42. Overall, the results support substantial training-free emotional editability, with mixed effects on transcription accuracy and a moderate reduction in reference-based speaker similarity.

For completeness, CosyVoice WER is 7.7287\% before alignment, 10.0759\% after alignment, and 10.6587\% after editing. The whole-pipeline increase is therefore 2.93 pp, including 2.35 pp from alignment. Its corresponding target-emotion probabilities are 3.6337\%, 12.4665\%, and 68.3269\%. This distinction separates alignment effects from the editing effects reported in Tables~\ref{tab:q1_probability_detail} and~\ref{tab:q1_wer_detail}.

\subsection{Experiment Details for Q2}
\label{app:editing_timing_q2}

\textbf{Experimental setup.}
We follow the diagnostic setup in Appendix~\ref{app:editability_diagnostic_q1} unless otherwise stated.
All trajectory interventions use the same 80 cases and model-specific inference settings.
Each intervention is evaluated with three editing-noise repeats while keeping the source synthesis fixed.
The reported $t$ denotes the model's internal integration step; the native solver grids are nonuniform.

\textbf{Trajectory-specific measures.}
We reuse the target-emotion probability change and $\Delta$WER defined in Appendix~\ref{app:editability_diagnostic_q1}.
For delayed-start experiments, we additionally report \emph{retained emotion gain}, defined as the mean emotion gain relative to full-step editing.
To quantify temporal disruption, we measure \emph{onset drift} as the absolute difference between the detected source and edited speech onsets.
Onset is detected using 20-ms RMS windows with a 10-ms hop and a threshold of 5\% of peak RMS.

\begin{table}[t]
    \centering
    \footnotesize
    \caption{Early-stop editing results. Editing begins at the first editing step and terminates after the indicated prefix. Results are averaged over 80 cases and three editing-noise repeats.}
    \label{tab:q2_early_stop}
    \scriptsize
    \begin{tabular}{llrrr}
        \toprule
        Model & Editing prefix & Emotion gain (pp) & Onset drift (ms) & $\Delta$WER (pp) \\
        \midrule
        F5-TTS & First 4 of 32 steps & 0.015 & 0.38 & +0.21 \\
        F5-TTS & All 32 steps & 53.78 & 308.33 & $-0.92$ \\
        \midrule
        CosyVoice 2 & First 3 of 10 steps & 3.43 & 22.21 & +2.83 \\
        CosyVoice 2 & First 6 of 10 steps & 30.37 & 87.83 & +88.88 \\
        CosyVoice 2 & All 10 steps & 56.09 & 245.42 & +1.40 \\
        \bottomrule
    \end{tabular}
\end{table}

\begin{table}[!htbp]
    \centering
    \footnotesize
    \caption{Selected delayed starts, averaged over cases and editing-noise repeats.
    Gain retention is relative to full-step editing; onset drift and $\Delta$WER are relative
    to the decoded, length-aligned source.}
    \label{tab:q2_delayed_start}
    \scriptsize
    \begin{tabular}{llrrr}
        \toprule
        Model & Editing start & Gain retained (\%) & Onset drift (ms) & $\Delta$WER (pp) \\
        \midrule
        F5-TTS & Full-step & 100.0 & 308.33 & $-0.92$ \\
        & Skip 4 steps & 99.5 & 97.33 & $-4.95$ \\
        \midrule
        CosyVoice 2 & Full-step & 100.0 & 245.42 & $+1.40$ \\
        & Skip 1 step & 100.3 & 260.29 & $+0.67$ \\
        & Skip 3 steps & 51.0 & 44.25 & $+55.75$ \\
        \bottomrule
    \end{tabular}
\end{table}

\begin{figure}[t]
    \centering
    \includegraphics[width=\textwidth]{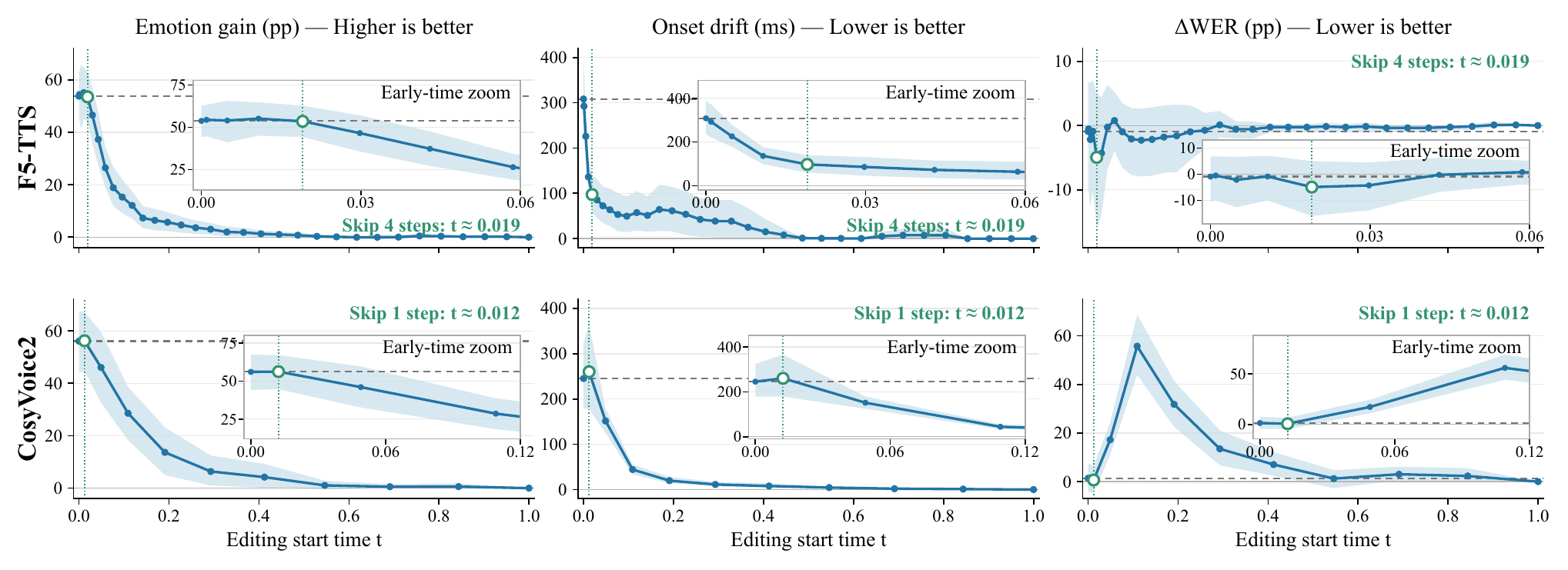}
    \caption{Complete delayed-start scans for F5-TTS and CosyVoice 2.
    Columns show emotion gain (pp), onset drift (ms), and $\Delta$WER (pp), measured relative
    to the decoded, length-aligned source.
    Bands indicate pointwise 95\% speaker-bootstrap intervals.
    The horizontal axis is the ODE step.}
    \label{fig:q2_start_later}
\end{figure}

\textbf{Delayed-start analysis.}
We first test whether the earliest editing updates are necessary.
Specifically, we hold the editing state unchanged for the first $k$ integration steps and apply the editing update at every remaining step.
We sweep all possible starting steps on the native schedules of F5-TTS (32 steps) and CosyVoice 2 (10 steps), including full-step editing and no editing.
Noise is shared across different start schedules at each corresponding integration step.

Table~\ref{tab:q2_delayed_start} reports representative operating points discussed in the main text, while Fig.~\ref{fig:q2_start_later} shows the complete scans.
F5-TTS exhibits a short early interval in which onset drift can be substantially reduced while retaining all of the full-step emotion gain.
CosyVoice 2 shows no similarly favorable delayed start: preserving the emotion gain provides little timing benefit, whereas stronger delay rapidly sacrifices editing effectiveness and linguistic fidelity.

\begin{figure}[!htbp]
    \centering
    \includegraphics[width=\textwidth]{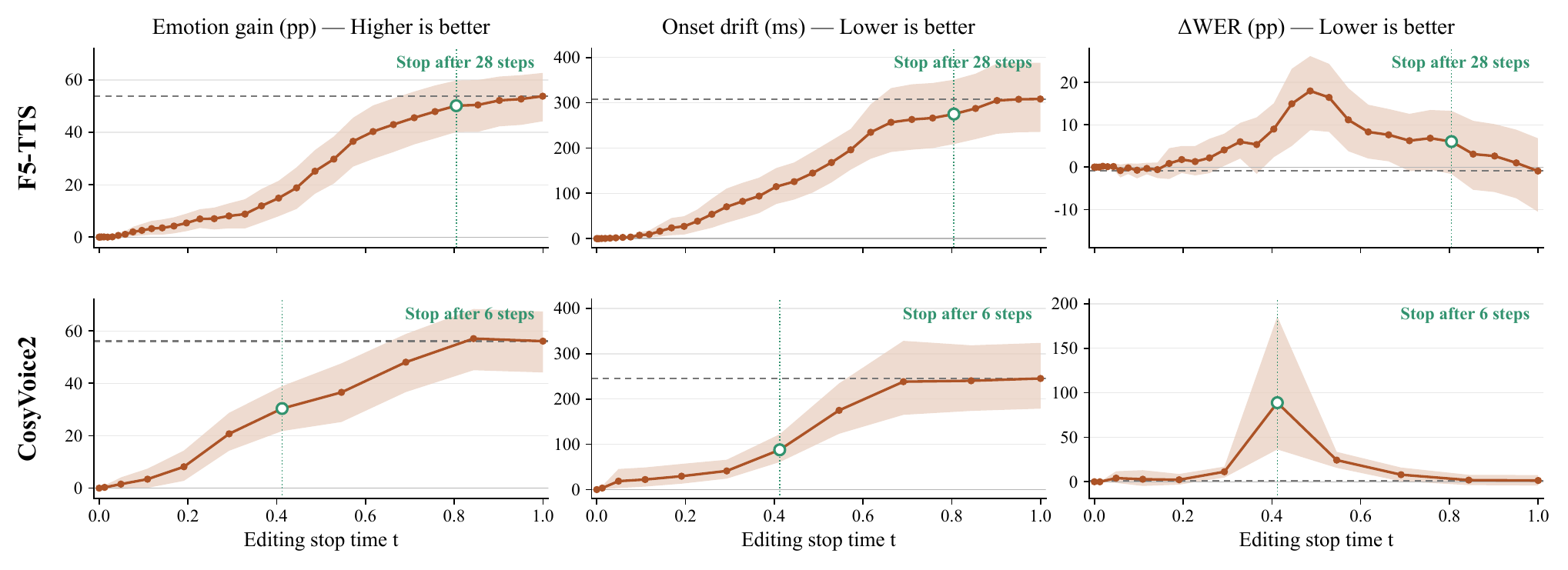}
    \caption{Early-stop scans for F5-TTS and CosyVoice 2. Columns report emotion gain, onset drift, and $\Delta$WER. Bands show 95\% speaker-bootstrap intervals; dashed lines indicate full-step means.}
    \label{fig:q2_stop_earlier}
\end{figure}

\textbf{Early-stop analysis.}
The delayed-start experiment tests whether early updates are \emph{necessary}.
Editing now begins at the first integration step and terminates after a selected prefix, after which the current state is frozen and decoded without subsequent editing updates.
Table~\ref{tab:q2_early_stop_full} reports representative results, and Fig.~\ref{fig:q2_stop_earlier} shows the complete scan.

Early prefixes produce little emotion change compared with full-step editing.
Moreover, intermediate states can have substantially worse transcription than the final state, showing that later updates may not only complete the emotion transfer but also recover intermediate content errors.

\begin{table}[!htbp]
    \centering
    \footnotesize
    \caption{Early-stop analysis.
    Editing starts at the first step and stops after the indicated prefix; values are means over
    80 cases and three editing-noise repeats.}
    \label{tab:q2_early_stop_full}
    \scriptsize
    \begin{tabular}{llrrrr}
        \toprule
        Model & Prefix & $t$ & Emotion gain (pp) & Onset drift (ms) & $\Delta$WER (pp) \\
        \midrule
        F5-TTS & First 4 / 32 & 0.019 & 0.015 & 0.38 & +0.21 \\
        F5-TTS & First 14 / 32 & 0.227 & 6.96 & 38.54 & +1.31 \\
        F5-TTS & First 28 / 32 & 0.805 & 50.15 & 275.04 & +6.03 \\
        F5-TTS & All 32 / 32 & 1.000 & 53.78 & 308.33 & $-0.92$ \\
        \midrule
        CosyVoice 2 & First 3 / 10 & 0.109 & 3.43 & 22.21 & +2.83 \\
        CosyVoice 2 & First 6 / 10 & 0.412 & 30.37 & 87.83 & +88.88 \\
        CosyVoice 2 & First 9 / 10 & 0.844 & 57.06 & 240.38 & +1.84 \\
        CosyVoice 2 & All 10 / 10 & 1.000 & 56.09 & 245.42 & +1.40 \\
        \bottomrule
    \end{tabular}
\end{table}

\begin{table}[t]
    \centering
    \footnotesize
    \caption{Block ablations.
    Values are emotion gain (pp), onset drift (ms), and $\Delta$WER (pp).}
    \label{tab:q2_block_ablation}
    \scriptsize

    \begin{tabular}{llrrrrrr}
        \toprule
        Model & Block & \multicolumn{3}{c}{Only block} & \multicolumn{3}{c}{Skip block} \\
        & & Gain & Drift & $\Delta$WER & Gain & Drift & $\Delta$WER \\
        \midrule
        F5-TTS & 1 & 6.96 & 38.54 & 1.31 & 4.66 & 53.71 & $-0.97$ \\
        & 2 & 1.17 & 14.17 & 0.60 & 36.00 & 205.38 & 13.88 \\
        & 3 & 0.72 & 0.50 & 0.25 & 40.27 & 253.75 & 8.10 \\
        & 4 & 0.23 & 2.62 & $-0.25$ & 49.26 & 267.79 & 4.95 \\
        & 5 & 0.36 & 7.38 & $-0.24$ & 50.15 & 275.04 & 6.03 \\
        \midrule
        CosyVoice 2 & 1 & 20.73 & 41.21 & 11.23 & 6.37 & 11.25 & 13.49 \\
        & 2 & $-0.27$ & 5.12 & 3.20 & 55.67 & 243.58 & 3.36 \\
        & 3 & 1.37 & 5.83 & 2.06 & 47.67 & 216.46 & 10.74 \\
        & 4 & 1.24 & 1.58 & 1.35 & 56.85 & 238.75 & 1.39 \\
        & 5 & 0.59 & 1.08 & 2.32 & 57.06 & 240.38 & 1.84 \\
        \bottomrule
    \end{tabular}
\end{table}

\begin{figure}[!htbp]
    \centering
    \includegraphics[width=\textwidth]{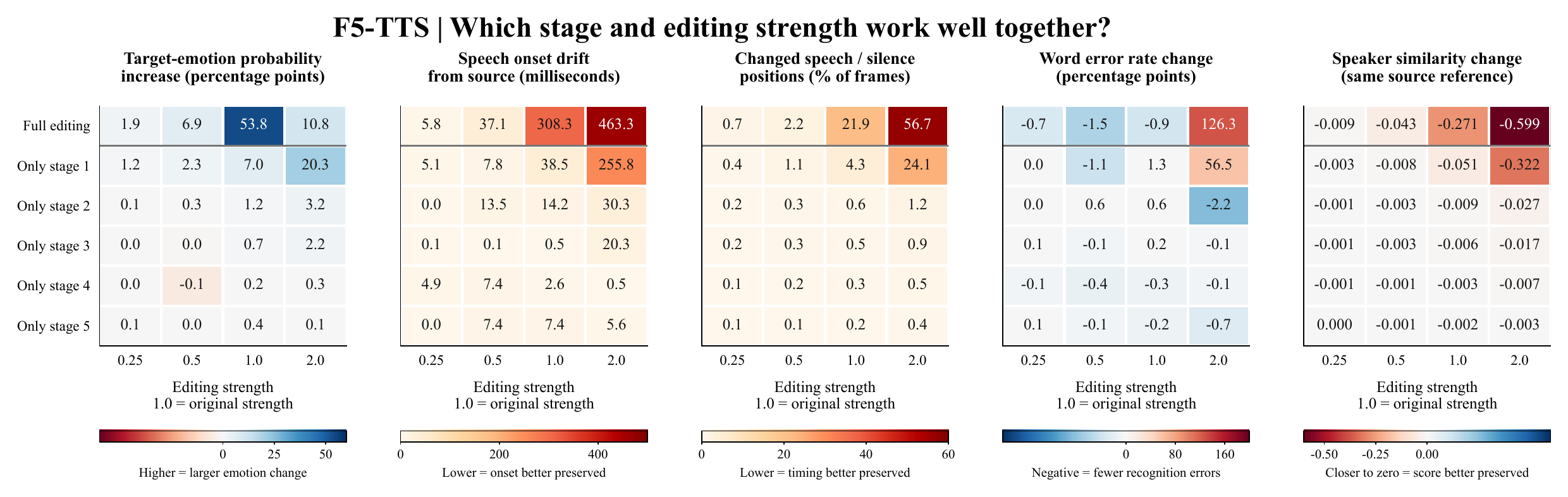}
    \caption{F5-TTS block and editing-strength ablations.}
    \label{fig:q2_f5_block_dose}
\end{figure}

\begin{figure}[!htbp]
    \centering
    \includegraphics[width=\textwidth]{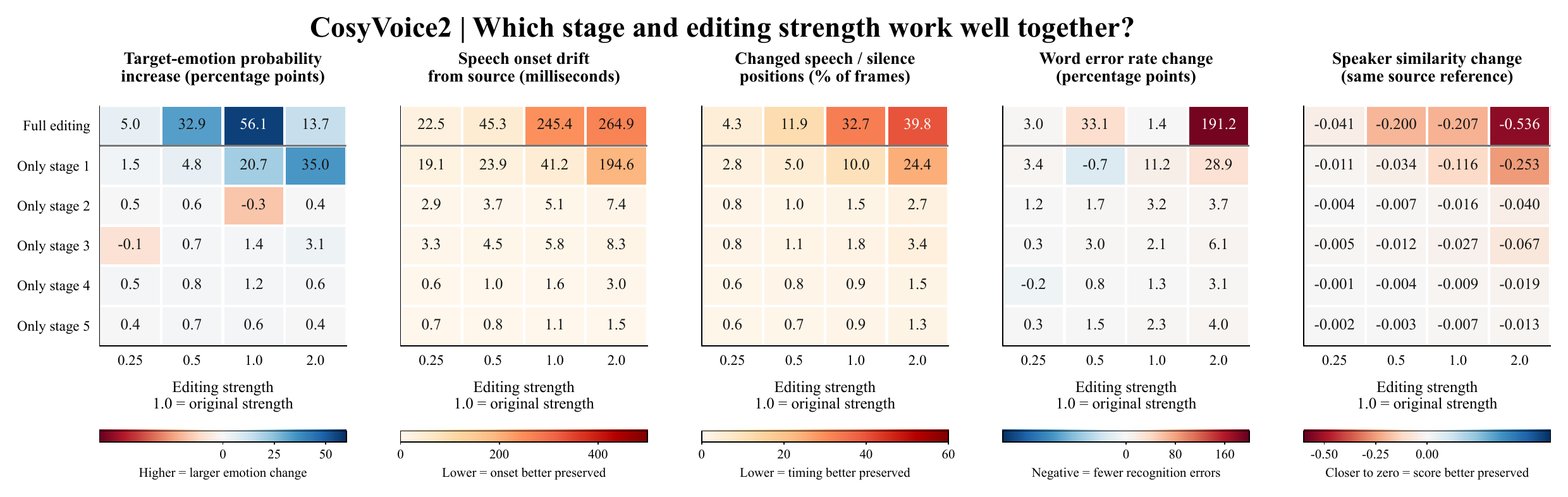}
    \caption{CosyVoice 2 block and editing-strength ablations.}
    \label{fig:q2_cosy_block_dose}
\end{figure}

\textbf{Block-wise and strength analysis.}
The start- and stop-time scans show that editing effects vary strongly along the trajectory, but do not establish whether individual stages contribute independently.
We therefore divide the trajectory into five stages and perform two complementary interventions.
``Only block $j$'' applies editing updates only within block $j$, whereas ``skip block $j$'' removes that block from an otherwise complete editing trajectory.

As shown in Table~\ref{tab:q2_block_ablation}, no individual block reproduces the emotion gain of full-step editing.
Conversely, removing a block from an otherwise complete trajectory can substantially change emotion gain, onset drift, or transcription accuracy.
The contribution of a stage therefore depends on the state produced by preceding updates, supporting a state-dependent trajectory interpretation rather than an additive assignment of independent roles to individual stages.

We further examine how editing strength interacts with trajectory position. Figs.~\ref{fig:q2_f5_block_dose} and~\ref{fig:q2_cosy_block_dose} (rows denote the full trajectory or individual blocks; columns denote editing strengths.
Cell values are means with metric-specific color scales) show that increasing the strength does not consistently improve emotion transfer. In some stages, stronger updates even weaken the emotion change while substantially increasing onset drift or transcription errors. Here, stages are defined by solver-time boundaries and therefore contain different numbers of integration steps, while the strength directly scales the editing update at each step. These results show that the appropriate editing strength depends on where along the trajectory it is applied.

\subsection{Experiment Details for Q3}
\label{app:q3_attributes}

\textbf{Experimental Setup.}
We reuse the source cases, full-trajectory editor, and emotion evaluator from Appendix~\ref{app:editability_diagnostic_q1}. Each Neutral source is paired with the next speaker in dataset-ID order within the same language, with wraparound, requesting Happy, Angry, Sad, or Surprise editing targets (80 edits per model).
Source and target references share the reference transcript; the synthesis text is unchanged.
Editing uses seed 42.
We keep the source mel spectrogram already length-aligned in Q1 fixed; the newly generated target mel segment is linearly interpolated to this fixed length and decoded for pitch comparison.
CosyVoice 2 retains the complete target condition, including target-generated speech tokens and flow-conditioning features in the original paper.

\textbf{Identity and pitch measures.}
For identity evaluation, each speaker has ten independent real enrollment recordings: two texts, each in five emotions, disjoint from all conditioning-reference and synthesis texts.
We unit-normalize their ECAPA-TDNN embeddings, average them, and normalize the mean to obtain a speaker centroid.
We measure identity changes relative to both the source and target centroids. For each centroid, we compute the cosine similarity before and after editing and report the edited-minus-source difference.

For pitch, we estimate F0 using the probabilistic YIN (pYIN) algorithm \citep{matthias2014pYIN} from 16-kHz mono audio, with a 1024-sample window, a 160-sample hop, and a 50-600 Hz search range.
Finite estimates are converted to semitones as $12\log_2(F0/100\,\mathrm{Hz})$.
For the median or 90th-minus-10th percentile range $f$, we report $|f(b)-f(t)|-|f(a)-f(t)|$, where $b$, $a$, and $t$ are the aligned source, edited output, and aligned target synthesis.
Positive values mean closer target pitch statistics.
Both models have 79/80 valid pitch comparisons, covering all 20 source speakers.

\textbf{Results.}
Table~\ref{tab:q3_attributes} reports the mean results across source speakers. We estimate uncertainty using 95\% confidence intervals from 10,000 source-speaker bootstrap resamples (seed 42), conditional on the fixed speaker pairs and generation seed and without correction for multiple comparisons. Overall, speaker identity, emotion, and pitch all shift toward the target.

\section{Additional Details of SEmoEditBench}
\label{app:benchmark}

\subsection{Task Construction}
\label{app:benchmark_tasks}

Table~\ref{tab:benchmark_splits} summarizes the nine benchmark splits. The same-dataset same-speaker split uses a paired target recording from the same speaker reading the same text, isolating emotion from speaker and content. The same-dataset cross-speaker split keeps the source corpus but uses a different speaker as the target reference. The cross-dataset cross-speaker split is drawn from a corpus different from the source corpus, with a different speaker. Every system receives the source waveform, its ground-truth transcript, the requested target emotion (and target intensity for intensity control), and a target reference utterance with different linguistic content. The paired target waveform is reserved exclusively for evaluation and must not be used to generate the edit. Sampling is deterministic and speaker-aware, and source or paired-target waveforms are never reused as target references.

\begin{table}[t]
\centering
\footnotesize
\caption{SEmoEditBench task composition. Each task is evaluated under same-dataset same-speaker, same-dataset cross-speaker, and cross-dataset cross-speaker settings.}
\label{tab:benchmark_splits}
\scriptsize
\begin{tabularx}{\textwidth}{@{}lllrX@{}}
\toprule
Task & Setting & Corpus & Cases & Construction \\
\midrule
Replacement & Same dataset, same speaker & ESD, IEMOCAP & 120 & 80 ESD and 40 IEMOCAP cases; sources and targets balanced across Neutral, Happy, Sad, Angry, and Surprise; Chinese and English in ESD. \\
Replacement & Same dataset, cross speaker & ESD, IEMOCAP & 120 & Same source cases as the same-speaker split, but the target reference comes from a different speaker of the same corpus. \\
Replacement & Cross dataset, cross speaker & ESD, IEMOCAP & 80 & Sources from one corpus and target references from the other; all English. \\
Erasure & Same dataset, same speaker & ESD, IEMOCAP & 56 & 32 ESD and 24 IEMOCAP cases; source emotions Happy, Sad, Angry, and Surprise, all mapped to Neutral. \\
Erasure & Same dataset, cross speaker & ESD, IEMOCAP & 56 & Same source cases, target reference from a different speaker of the same corpus. \\
Erasure & Cross dataset, cross speaker & ESD, IEMOCAP & 40 & Sources from one corpus and target references from the other; all English. \\
Intensity & Same dataset, same speaker & RAVDESS, CREMA-D & 44 & 20 RAVDESS and 24 CREMA-D cases; Neutral sources paired with same-speaker same-text emotional targets at five intensities. \\
Intensity & Same dataset, cross speaker & RAVDESS, CREMA-D & 44 & Same source cases, target reference from a different speaker of the same corpus. \\
Intensity & Cross dataset, cross speaker & RAVDESS, CREMA-D & 40 & Sources from one corpus and target references from the other; all English. \\
\bottomrule
\end{tabularx}
\end{table}

IEMOCAP sources are restricted to utterances between 2 and 10 seconds with categorical annotation agreement of at least 0.6. For ESD, IEMOCAP, and RAVDESS, the target reference matches the source speaker but uses different text. Because CREMA-D does not provide suitable same-speaker references for this protocol, its references are synthesized with IndexTTS2 \citep{zhou2026indextts2}: the neutral source provides speaker identity, a separate recording provides the requested emotion and intensity, and a synthesis transcript distinct from the source provides the reference content. The benchmark fixes all manifests, case identifiers, and references across evaluated systems. All intensity split summaries are averaged over the same 90-case common subset (Neutral source; Angry, Happy, Sad, or Surprise target; 30 cases per split) so that systems are directly comparable.

\subsection{Metric Definitions}
\label{app:benchmark_metrics}

We evaluate each edit along three axes: editing success, preservation of non-target attributes, and perceptual quality.
All objective metrics are computed per case and macro-averaged within each split using identical model configurations across systems; expected case count, failed case count, and failure rate are reported alongside the metric averages.
Table~\ref{tab:notation} lists the symbols used throughout the metric definitions.
Table~\ref{tab:benchmark_metrics} provides a compact overview of all metrics, their optimization directions, and task-level applicability; the subsections below define each metric in detail.

\begin{table}[!htbp]
\centering
\footnotesize
\caption{Notation used in the metric definitions.}
\label{tab:notation}
\scriptsize
\begin{tabularx}{\textwidth}{@{}>{\raggedright\arraybackslash}p{0.18\textwidth}>{\raggedright\arraybackslash}X@{}}
\toprule
Symbol & Meaning \\
\midrule
$x_s$, $x_e$, $x_t$ & Source, edited, and paired-target waveforms, respectively \\
$e_{\mathrm{src}}$, $e_{\mathrm{tgt}}$ & Source and target emotion labels \\
$y$ & Manifest transcript (ground-truth text) \\
$P(e\mid x)$ & emotion2vec+ Large posterior probability for emotion $e$ given waveform $x$ \\
$f_E(x)$ & emotion2vec+ Large embedding of waveform $x$ \\
$A(x)$ & ASR transcript of waveform $x$ produced by Whisper-large-v3 \\
$f_S(x)$ & ECAPA-TDNN speaker embedding of waveform $x$ \\
$Q(x)$ & UTMOSv2 quality prediction for waveform $x$ \\
$x_e^{(\alpha)}$ & Edited output at intensity strength $\alpha$ \\
$\alpha_1<\cdots<\alpha_5$ & Five intensity strengths $\{0, 0.25, 0.5, 0.75, 1\}$ \\
$z_s, z_t, z_i$ & emotion2vec embeddings $f_E(x_s)$, $f_E(x_t)$, $f_E(x_e^{(\alpha_i)})$ \\
$a_i$ & Relative progress of output $i$ along the source--target direction: $(z_i-z_s)^\top(z_t-z_s)/\lVert z_t-z_s\rVert^2$ \\
\bottomrule
\end{tabularx}
\end{table}

\begin{table}[t]
\centering
\footnotesize
\begin{threeparttable}
\caption{Metrics used in SEmoEditBench. Upward and downward arrows indicate whether higher or lower values are preferred.}
\label{tab:benchmark_metrics}
\scriptsize
\begin{tabularx}{\textwidth}{@{}>{\raggedright\arraybackslash}p{0.27\textwidth}>{\raggedright\arraybackslash}X>{\raggedright\arraybackslash}p{0.18\textwidth}@{}}
\toprule
Metric & Definition & Applicability \\
\midrule
\multicolumn{3}{@{}l}{\textbf{Editing success}} \\
\addlinespace[2pt]
Target emotion probability (TEP) $\uparrow$ & $P(e_{\mathrm{tgt}}\mid x_e)$ & Replacement \\
Neutral probability (NP) $\uparrow$ & $P(\mathrm{Neutral}\mid x_e)$ & Erasure \\
Source emotion suppression (SES) $\uparrow$ & $P(e_{\mathrm{src}}\mid x_s)-P(e_{\mathrm{src}}\mid x_e)$ & Replacement, erasure \\
Embedding-based effective intensity control (EIC-Emb) $\uparrow$ & Mean over all pairs $i<j$ of $\operatorname{sgn}(a_j-a_i)\cdot\min(1,\max(0,|a_j-a_i|/\tau))$, $\tau=1$ & Intensity control (paired target) \\
Emotion similarity (E-SIM) $\uparrow$ & $\cos\!\left(f_E(x_e),f_E(x_t)\right)$ & Splits with paired targets \\
Directional editing score (DES) $\uparrow$ & $\cos\!\left(f_E(x_e)-f_E(x_s),f_E(x_t)-f_E(x_s)\right)$ & Splits with paired targets \\
\midrule
\multicolumn{3}{@{}l}{\textbf{Preservation}} \\
\addlinespace[2pt]
Relative word error rate ($\Delta$WER) $\downarrow$ & $\mathrm{WER}(x_e)-\mathrm{WER}(x_s)$ against the same transcript; effectively $\Delta$CER for Chinese & All tasks \\
Speaker similarity (S-SIM) $\uparrow$ & ECAPA-TDNN cosine verification score between $x_s$ and $x_e$ & All tasks \\
\midrule
\multicolumn{3}{@{}l}{\textbf{Quality}} \\
\addlinespace[2pt]
UTMOS $\uparrow$ & UTMOSv2 prediction for $x_e$ & All tasks \\
\midrule
\multicolumn{3}{@{}l}{\textbf{Subjective evaluation}} \\
\addlinespace[2pt]
Speaker-similarity MOS (SS-MOS) $\uparrow$ & Perceived speaker similarity between the source and edited utterances & 20 sampled case groups \\
Emotion similarity (ES-MOS) $\uparrow$ & Perceived emotion similarity between the edited utterance and requested target emotion & 20 sampled case groups \\
Naturalness MOS (N-MOS) $\uparrow$ & Perceived naturalness of the edited utterance & 20 sampled case groups \\
\bottomrule
\end{tabularx}
\begin{tablenotes}[flushleft]
\scriptsize
\item For DES, a zero edited displacement receives a score of zero, whereas a zero target displacement is invalid. E-SIM and DES are reported for all three replacement and erasure splits, each of which provides an evaluation-only paired target recording. EIC-Emb uses the corresponding paired target recordings for all three intensity splits, never the generation-time target reference.
\end{tablenotes}
\end{threeparttable}
\end{table}

\subsubsection{Editing Success}
\label{app:benchmark_metrics_success}

Editing-success metrics measure whether the edited waveform expresses the requested target emotion (or, for erasure, neutral) while moving away from the source emotion. They are computed from emotion2vec+ Large posteriors and embeddings.

\textbf{Target emotion probability (TEP).}
TEP measures the degree to which the edited waveform is recognized as the requested target emotion:
\begin{equation}
\mathrm{TEP} = P(e_{\mathrm{tgt}} \mid x_e).
\end{equation}
\emph{Measurement.} The emotion2vec+ Large model produces a posterior distribution over emotion categories from the edited waveform; the probability mass assigned to $e_{\mathrm{tgt}}$ is reported. \emph{Data source.} Edited waveform $x_e$. \emph{Evaluation criteria.} Higher values indicate stronger target-emotion expression; reported for the replacement task.

\textbf{Neutral probability (NP).}
For emotion erasure, the requested output is emotionally neutral. NP measures the probability that the edited waveform is classified as neutral:
\begin{equation}
\mathrm{NP} = P(\mathrm{neutral} \mid x_e).
\end{equation}
\emph{Measurement.} Same as TEP but extracting the posterior mass for the neutral class. \emph{Data source.} Edited waveform $x_e$. \emph{Evaluation criteria.} Higher values indicate more complete erasure; reported for the erasure task.

\textbf{Source emotion suppression (SES).}
SES quantifies the reduction of the source emotion from the source to the edited waveform:
\begin{equation}
\mathrm{SES} = P(e_{\mathrm{src}} \mid x_s) - P(e_{\mathrm{src}} \mid x_e).
\end{equation}
\emph{Measurement.} The emotion2vec posterior for the source emotion $e_{\mathrm{src}}$ is computed for both the source and edited waveforms, and their difference is taken. \emph{Data sources.} Source waveform $x_s$ and edited waveform $x_e$. \emph{Evaluation criteria.} Higher values indicate stronger suppression of the source emotion; reported for both replacement and erasure.

\textbf{Embedding-based effective intensity control (EIC-Emb).}
EIC-Emb is the primary metric used to evaluate intensity control in the main benchmark. For each case, the system produces five edited outputs $x_e^{(\alpha_i)}$ at increasing strengths $\alpha_1<\cdots<\alpha_5$. Each output embedding is projected onto the source--target direction via the relative progress
\begin{equation} \label{eq:eic_proj}
a_i = \frac{(z_i - z_s)^\top (z_t - z_s)}{\lVert z_t - z_s \rVert^2},
\end{equation}
where $a_i=0$ means no progress beyond the source, $a_i=1$ reaches the target's projection, negative values move away from the target, and values above $1$ overshoot. EIC-Emb then applies a pairwise effective-control transform: for every pair $i<j$, the signed progress difference $d=a_j-a_i$ is mapped to
\begin{equation} \label{eq:eic_sign}
\operatorname{sgn}(d)\cdot\min\!\left(1,\max\!\left(0,\frac{|d|}{\tau}\right)\right), \qquad \tau=1,
\end{equation}
and the mean over all $\binom{5}{2}=10$ pairs is reported.
\emph{Measurement.} emotion2vec+ Large embeddings are computed for the source, the paired target, and all five edited outputs; the relative progress values $a_i$ are derived and the pairwise formula above is applied. \emph{Data sources.} Source waveform $x_s$, paired-target waveform $x_t$ (evaluation-only), and the five edited outputs $x_e^{(\alpha_i)}$. \emph{Evaluation criteria.} Higher values indicate more effective and monotonic control; a score of $1$ means every stronger edit produces strictly more progress toward the target with differences of at least $\tau$, $0$ means no consistent increase, and negative values indicate decreasing progress. EIC-Emb is reported for the intensity-control task on cases that have a paired target recording. Because $a_i$ is a signed projection rather than a cosine similarity, EIC-Emb preserves the magnitude of the emotional displacement and is anchored to the actual paired target recording.

\textbf{Emotion similarity (E-SIM).}
E-SIM measures the cosine similarity between the edited waveform's embedding and the paired target's embedding:
\begin{equation}
\mathrm{E\text{-}SIM} = \cos\!\left(f_E(x_e), f_E(x_t)\right),
\qquad
\cos(u,v)=\frac{u^\top v}{\lVert u\rVert_2\lVert v\rVert_2}.
\end{equation}
\emph{Measurement.} emotion2vec+ Large embeddings are computed for both waveforms and their cosine similarity is reported. \emph{Data sources.} Edited waveform $x_e$ and paired-target waveform $x_t$. \emph{Evaluation criteria.} Higher values indicate that the edit matches the emotional expression of the target; reported for splits with paired targets.

\textbf{Directional editing score (DES).}
DES measures whether the edit moves in the same direction as the target relative to the source:
\begin{equation}
d_e = f_E(x_e) - f_E(x_s), \qquad d_t = f_E(x_t) - f_E(x_s),
\qquad
\mathrm{DES} = \cos(d_e, d_t).
\end{equation}
A zero edited displacement $d_e=\mathbf{0}$ receives a score of zero, whereas a zero target displacement $d_t=\mathbf{0}$ is invalid. \emph{Measurement.} emotion2vec+ Large embeddings are computed for the source, edited, and paired-target waveforms; the displacement vectors are formed and their cosine similarity is reported. \emph{Data sources.} Source $x_s$, edited $x_e$, and paired-target $x_t$ waveforms. \emph{Evaluation criteria.} Higher values indicate that the edit travels along the correct source-to-target emotional direction; reported for splits with paired targets.

\subsubsection{Preservation}
\label{app:benchmark_metrics_preservation}

Preservation metrics verify that the edit modifies only the intended affective expression while keeping the linguistic content and speaker identity intact.

\textbf{Relative word error rate ($\Delta$WER).}
$\Delta$WER measures the change in transcription error from the source to the edited waveform against the same ground-truth transcript $y$:
\begin{equation}
\Delta\mathrm{WER} = \mathrm{WER}\left(y, A(x_e)\right) - \mathrm{WER}\left(y, A(x_s)\right).
\end{equation}
\emph{Measurement.} Both the source and edited waveforms are transcribed by Whisper-large-v3; WER is computed against the manifest transcript $y$ after text normalization. For Chinese, tokenization is character-level, making this effectively a character error rate ($\Delta$CER). \emph{Data sources.} Source waveform $x_s$, edited waveform $x_e$, and ground-truth transcript $y$. \emph{Evaluation criteria.} Lower values indicate better content preservation; $\Delta\mathrm{WER}=0$ means the edit introduced no transcription errors relative to the source. Reported for all tasks.

\textbf{Speaker similarity (S-SIM).}
S-SIM measures the cosine similarity between the speaker embeddings of the source and edited waveforms:
\begin{equation}
\mathrm{S\text{-}SIM} = \cos\!\left(f_S(x_s), f_S(x_e)\right).
\end{equation}
\emph{Measurement.} A pretrained ECAPA-TDNN speaker verifier extracts embeddings from both waveforms, and their cosine similarity is reported. \emph{Data sources.} Source waveform $x_s$ and edited waveform $x_e$. \emph{Evaluation criteria.} Higher values indicate better preservation of the source speaker identity; reported for all tasks.

\subsubsection{Quality}
\label{app:benchmark_metrics_quality}

\textbf{UTMOS.}
UTMOS estimates the perceptual naturalness of the edited waveform:
\begin{equation}
\mathrm{UTMOS} = Q(x_e).
\end{equation}
\emph{Measurement.} The pretrained UTMOSv2\footnote{\url{https://github.com/sarulab-speech/UTMOSv2}} predictor takes the edited waveform as input and outputs a quality score. \emph{Data source.} Edited waveform $x_e$. \emph{Evaluation criteria.} Higher values indicate better predicted audio quality; reported for all tasks.

\subsubsection{Subjective Evaluation}
\label{app:benchmark_metrics_subjective}

In addition to objective metrics, we uniformly sample 20 case groups from the benchmark and collect human judgments on a five-point scale ($1$--$5$) for three attributes.

\textbf{Speaker-similarity MOS (SS-MOS).}
\emph{Definition.} Perceived speaker similarity between the source and edited utterances. \emph{Measurement.} Raters compare the source and edited waveforms and score how well the speaker identity is preserved. \emph{Evaluation criteria.} Higher scores indicate better speaker preservation.

\textbf{Emotion similarity MOS (ES-MOS).}
\emph{Definition.} Perceived emotion similarity between the edited utterance and the requested target emotion. \emph{Measurement.} Raters are given the edited waveform together with the target emotion label (and the target reference when available) and score how closely the edit matches the intended emotion. \emph{Evaluation criteria.} Higher scores indicate closer emotional match to the target.

\textbf{Naturalness MOS (N-MOS).}
\emph{Definition.} Perceived naturalness of the edited utterance. \emph{Measurement.} Raters listen to the edited waveform and score its overall naturalness, including prosody, absence of artifacts, and fluency. \emph{Evaluation criteria.} Higher scores indicate more natural-sounding speech.

\newcommand{\idssfirst}[1]{\cellcolor{pink!60}\textbf{#1}}
\newcommand{\idsssecond}[1]{\cellcolor{pink!40}\textbf{#1}}
\newcommand{\idssthird}[1]{\cellcolor{pink!20}\textbf{#1}}
\newcommand{\idcsfirst}[1]{\cellcolor{green!40}\textbf{#1}}
\newcommand{\idcssecond}[1]{\cellcolor{green!30}\textbf{#1}}
\newcommand{\idcsthird}[1]{\cellcolor{green!20}\textbf{#1}}
\newcommand{\oodfirst}[1]{\cellcolor{blue!40}\textbf{#1}}
\newcommand{\oodsecond}[1]{\cellcolor{blue!30}\textbf{#1}}
\newcommand{\oodthird}[1]{\cellcolor{blue!20}\textbf{#1}}

\begin{table*}[t]
    \centering
    \caption{Complete emotion-replacement results under both in-distribution settings and the out-of-distribution setting.}
    \label{tab:complete_replacement_results}
    \scriptsize
    \renewcommand{\arraystretch}{0.8}
    \setlength{\tabcolsep}{4.3pt}
    \begin{tabular}{llllccccccc}
        \toprule
        Category & Method & Backbone & Setting & TEP $\uparrow$ & SES $\uparrow$ & E-SIM $\uparrow$ & DES $\uparrow$ & $\Delta$WER $\downarrow$ & S-SIM $\uparrow$ & UTMOS $\uparrow$ \\
        \midrule
        \multirow{9}{*}{\shortstack{Training-\\based}} & \multirow{3}{*}{Step-Audio-EditX} & \multirow{3}{*}{--} & ID-SS & 0.222 & 0.445 & 0.521 & 0.550 & \idssfirst{-0.048} & 0.567 & \idssfirst{3.091} \\
        & & & ID-CS & 0.196 & 0.420 & 0.519 & 0.543 & -0.021 & 0.561 & \idcsfirst{3.053} \\
        & & & OOD & 0.211 & 0.452 & 0.526 & 0.575 & -0.026 & 0.519 & \oodthird{3.142} \\
        \cmidrule(lr){2-11}
        & \multirow{3}{*}{Auk} & \multirow{3}{*}{--} & ID-SS & 0.259 & 0.372 & 0.539 & 0.541 & 0.073 & 0.631 & 2.639 \\
        & & & ID-CS & 0.259 & 0.372 & 0.539 & 0.541 & 0.073 & 0.631 & 2.670 \\
        & & & OOD & \oodsecond{0.312} & 0.353 & \oodsecond{0.586} & 0.580 & 0.103 & 0.580 & 2.753 \\
        \cmidrule(lr){2-11}
        & \multirow{3}{*}{dots.tts.edit} & \multirow{3}{*}{--} & ID-SS & 0.434 & 0.600 & 0.669 & 0.713 & \idssthird{-0.035} & 0.272 & 2.381 \\
        & & & ID-CS & \idcsthird{0.434} & 0.600 & \idcsthird{0.669} & \idcsthird{0.713} & -0.035 & 0.272 & 2.394 \\
        & & & OOD & \oodfirst{0.357} & 0.483 & \oodfirst{0.638} & \oodfirst{0.691} & \oodthird{-0.051} & 0.215 & 2.512 \\
        \midrule
        \multirow{15}{*}{\shortstack{Activation-\\steering}} & \multirow{3}{*}{CoCoEmo} & \multirow{3}{*}{CosyVoice~2} & ID-SS & 0.082 & 0.192 & 0.446 & 0.399 & -0.015 & \idssthird{0.715} & \idsssecond{3.061} \\
        & & & ID-CS & 0.133 & 0.189 & 0.447 & 0.364 & -0.014 & \idcsthird{0.717} & \idcssecond{3.046} \\
        & & & OOD & 0.136 & 0.217 & 0.431 & 0.470 & -0.039 & \oodthird{0.667} & \oodsecond{3.169} \\
        \cmidrule(lr){2-11}
        & \multirow{3}{*}{CoCoEmo} & \multirow{3}{*}{IndexTTS2} & ID-SS & 0.035 & 0.127 & 0.402 & 0.269 & -0.024 & \idsssecond{0.776} & 2.659 \\
        & & & ID-CS & 0.034 & 0.101 & 0.396 & 0.298 & -0.019 & \idcssecond{0.782} & 2.658 \\
        & & & OOD & 0.046 & 0.125 & 0.398 & 0.334 & -0.024 & \oodsecond{0.735} & 2.723 \\
        \cmidrule(lr){2-11}
        & \multirow{3}{*}{EmoSteer-TTS} & \multirow{3}{*}{F5-TTS} & ID-SS & 0.081 & 0.345 & 0.448 & 0.477 & 0.608 & 0.623 & 2.545 \\
        & & & ID-CS & 0.081 & 0.345 & 0.448 & 0.477 & 0.608 & 0.623 & 2.507 \\
        & & & OOD & 0.117 & 0.449 & 0.475 & 0.562 & 0.904 & 0.550 & 2.503 \\
        \cmidrule(lr){2-11}
        & \multirow{3}{*}{EmoSteer-TTS} & \multirow{3}{*}{CosyVoice~2} & ID-SS & 0.030 & 0.371 & 0.382 & 0.367 & -0.015 & 0.478 & 2.515 \\
        & & & ID-CS & 0.030 & 0.371 & 0.382 & 0.367 & -0.015 & 0.478 & 2.523 \\
        & & & OOD & 0.042 & 0.396 & 0.384 & 0.460 & -0.029 & 0.425 & 2.530 \\
        \cmidrule(lr){2-11}
        & \multirow{3}{*}{EmoSteer-TTS} & \multirow{3}{*}{IndexTTS2} & ID-SS & 0.025 & 0.054 & 0.377 & 0.226 & \idsssecond{-0.037} & \idssfirst{0.788} & 2.666 \\
        & & & ID-CS & 0.025 & 0.054 & 0.377 & 0.226 & \idcsthird{-0.037} & \idcsfirst{0.788} & 2.687 \\
        & & & OOD & 0.037 & 0.069 & 0.361 & 0.240 & \oodfirst{-0.060} & \oodfirst{0.758} & 2.707 \\
        \midrule
        \multirow{9}{*}{\textbf{SEmoEdit}} & \multirow{3}{*}{\textbf{Ours}} & \multirow{3}{*}{F5-TTS} & ID-SS & \idssthird{0.498} & \idssthird{0.684} & \idssthird{0.704} & \idssthird{0.753} & -0.026 & 0.372 & 2.177 \\
        & & & ID-CS & 0.432 & \idcsfirst{0.692} & 0.646 & 0.700 & \idcsfirst{-0.049} & 0.434 & 2.494 \\
        & & & OOD & 0.283 & \oodsecond{0.544} & 0.563 & \oodthird{0.629} & \oodsecond{-0.057} & 0.395 & 2.466 \\
        \cmidrule(lr){2-11}
        & \multirow{3}{*}{\textbf{Ours}} & \multirow{3}{*}{CosyVoice~2} & ID-SS & \idsssecond{0.554} & \idsssecond{0.694} & \idsssecond{0.776} & \idsssecond{0.806} & -0.025 & 0.379 & \idssthird{2.984} \\
        & & & ID-CS & \idcssecond{0.460} & \idcssecond{0.683} & \idcssecond{0.713} & \idcssecond{0.752} & \idcssecond{-0.045} & 0.438 & \idcsthird{2.925} \\
        & & & OOD & 0.268 & \oodthird{0.539} & 0.550 & \oodsecond{0.640} & -0.047 & 0.398 & \oodfirst{3.206} \\
        \cmidrule(lr){2-11}
        & \multirow{3}{*}{\textbf{Ours}} & \multirow{3}{*}{IndexTTS2} & ID-SS & \idssfirst{0.691} & \idssfirst{0.767} & \idssfirst{0.902} & \idssfirst{0.920} & -0.004 & 0.423 & 2.578 \\
        & & & ID-CS & \idcsfirst{0.516} & \idcsthird{0.671} & \idcsfirst{0.768} & \idcsfirst{0.775} & 0.000 & 0.468 & 2.686 \\
        & & & OOD & \oodthird{0.297} & \oodfirst{0.547} & \oodthird{0.576} & 0.585 & -0.046 & 0.436 & 2.930 \\
        \bottomrule
    \end{tabular}
\end{table*}

\begin{table*}[t]
    \centering
    \caption{Complete emotion-erasure results under both in-distribution settings and the out-of-distribution setting. CoCoEmo is omitted because it does not support emotion erasure.}
    \label{tab:complete_erasure_results}
    \scriptsize
    \renewcommand{\arraystretch}{0.8}
    \setlength{\tabcolsep}{4.3pt}
    \begin{tabular}{llllccccccc}
        \toprule
        Category & Method & Backbone & Setting & NP $\uparrow$ & SES $\uparrow$ & E-SIM $\uparrow$ & DES $\uparrow$ & $\Delta$WER $\downarrow$ & S-SIM $\uparrow$ & UTMOS $\uparrow$ \\
        \midrule
        \multirow{9}{*}{\shortstack{Training-\\based}} & \multirow{3}{*}{Step-Audio-EditX} & \multirow{3}{*}{--} & ID-SS & 0.235 & 0.388 & 0.566 & 0.655 & \idsssecond{-0.064} & 0.577 & \idssfirst{3.220} \\
        & & & ID-CS & 0.347 & 0.488 & 0.606 & 0.688 & -0.028 & 0.586 & \idcssecond{3.133} \\
        & & & OOD & 0.269 & 0.465 & \oodthird{0.746} & \oodsecond{0.818} & \oodthird{-0.070} & \oodthird{0.561} & \oodsecond{3.197} \\
        \cmidrule(lr){2-11}
        & \multirow{3}{*}{Auk} & \multirow{3}{*}{--} & ID-SS & 0.229 & 0.311 & 0.528 & 0.582 & 0.014 & \idsssecond{0.676} & \idssthird{2.859} \\
        & & & ID-CS & 0.229 & 0.311 & 0.528 & 0.582 & 0.014 & \idcssecond{0.676} & \idcsthird{2.881} \\
        & & & OOD & 0.272 & 0.371 & 0.591 & 0.585 & 0.033 & \oodsecond{0.651} & 2.959 \\
        \cmidrule(lr){2-11}
        & \multirow{3}{*}{dots.tts.edit} & \multirow{3}{*}{--} & ID-SS & 0.355 & 0.549 & 0.674 & 0.766 & \idssfirst{-0.065} & 0.304 & 2.685 \\
        & & & ID-CS & 0.355 & 0.549 & 0.674 & 0.766 & \idcssecond{-0.065} & 0.304 & 2.675 \\
        & & & OOD & \oodthird{0.350} & \oodthird{0.501} & 0.745 & 0.790 & \oodthird{-0.070} & 0.254 & 2.752 \\
        \midrule
        \multirow{9}{*}{\shortstack{Activation-\\steering}} & \multirow{3}{*}{EmoSteer-TTS} & \multirow{3}{*}{F5-TTS} & ID-SS & 0.202 & 0.393 & 0.616 & 0.736 & 0.030 & \idssthird{0.598} & 2.593 \\
        & & & ID-CS & 0.202 & 0.393 & 0.616 & 0.736 & 0.030 & \idcsthird{0.598} & 2.575 \\
        & & & OOD & 0.258 & 0.451 & 0.731 & \oodthird{0.812} & 0.030 & 0.549 & 2.643 \\
        \cmidrule(lr){2-11}
        & \multirow{3}{*}{EmoSteer-TTS} & \multirow{3}{*}{CosyVoice~2} & ID-SS & 0.078 & 0.174 & 0.505 & 0.596 & -0.012 & 0.548 & 2.490 \\
        & & & ID-CS & 0.078 & 0.174 & 0.505 & 0.596 & -0.012 & 0.548 & 2.491 \\
        & & & OOD & 0.084 & 0.167 & 0.605 & 0.682 & -0.007 & 0.510 & 2.519 \\
        \cmidrule(lr){2-11}
        & \multirow{3}{*}{EmoSteer-TTS} & \multirow{3}{*}{IndexTTS2} & ID-SS & 0.038 & 0.070 & 0.377 & 0.298 & -0.041 & \idssfirst{0.808} & 2.836 \\
        & & & ID-CS & 0.038 & 0.070 & 0.377 & 0.298 & -0.041 & \idcsfirst{0.808} & 2.845 \\
        & & & OOD & 0.053 & 0.099 & 0.452 & 0.317 & -0.045 & \oodfirst{0.798} & 2.853 \\
        \midrule
        \multirow{9}{*}{\textbf{SEmoEdit}} & \multirow{3}{*}{\textbf{Ours}} & \multirow{3}{*}{F5-TTS} & ID-SS & \idssthird{0.573} & \idssthird{0.700} & \idssthird{0.821} & \idssthird{0.873} & \idssthird{-0.058} & 0.359 & 2.597 \\
        & & & ID-CS & \idcsfirst{0.592} & \idcssecond{0.670} & \idcssecond{0.802} & \idcsfirst{0.857} & \idcsfirst{-0.081} & 0.409 & 2.706 \\
        & & & OOD & \oodsecond{0.408} & \oodfirst{0.557} & \oodsecond{0.758} & \oodfirst{0.820} & \oodfirst{-0.098} & 0.399 & 2.665 \\
        \cmidrule(lr){2-11}
        & \multirow{3}{*}{\textbf{Ours}} & \multirow{3}{*}{CosyVoice~2} & ID-SS & \idssfirst{0.704} & \idssfirst{0.754} & \idsssecond{0.907} & \idssfirst{0.938} & -0.050 & 0.399 & \idsssecond{3.144} \\
        & & & ID-CS & \idcssecond{0.580} & \idcsfirst{0.687} & \idcsfirst{0.839} & \idcssecond{0.854} & \idcsthird{-0.047} & 0.443 & \idcsfirst{3.166} \\
        & & & OOD & \oodfirst{0.418} & \oodsecond{0.532} & \oodfirst{0.772} & 0.777 & \oodsecond{-0.078} & 0.431 & \oodfirst{3.303} \\
        \cmidrule(lr){2-11}
        & \multirow{3}{*}{\textbf{Ours}} & \multirow{3}{*}{IndexTTS2} & ID-SS & \idsssecond{0.682} & \idsssecond{0.741} & \idssfirst{0.910} & \idsssecond{0.934} & -0.030 & 0.424 & 2.774 \\
        & & & ID-CS & \idcsthird{0.576} & \idcsthird{0.634} & \idcsthird{0.798} & \idcsthird{0.806} & -0.032 & 0.470 & 2.783 \\
        & & & OOD & 0.284 & 0.385 & 0.692 & 0.729 & -0.064 & 0.460 & \oodthird{3.050} \\
        \bottomrule
    \end{tabular}
\end{table*}

\begin{table*}[t]
    \centering
    \caption{Complete emotion-intensity-control results under both in-distribution settings and the out-of-distribution setting. Training-based baselines are omitted because they do not support continuous intensity control.}
    \label{tab:complete_intensity_results}
    \scriptsize
    \renewcommand{\arraystretch}{0.8}
    \begin{tabular}{llllcccc}
        \toprule
        Category & Method & Backbone & Setting & EIC-Emb $\uparrow$ & $\Delta$WER$_{\alpha=1}$ $\downarrow$ & S-SIM$_{\alpha=1}$ $\uparrow$ & UTMOS$_{\alpha=1}$ $\uparrow$ \\
        \midrule
        \multirow{15}{*}{\shortstack{Activation-\\steering}} & \multirow{3}{*}{CoCoEmo} & \multirow{3}{*}{CosyVoice~2} & ID-SS & 0.041 & \idssthird{0.044} & \idssthird{0.621} & \idssfirst{3.234} \\
        & & & ID-CS & 0.023 & \idcssecond{0.006} & \idcsthird{0.648} & \idcsfirst{3.206} \\
        & & & OOD & 0.057 & \oodthird{0.044} & \oodthird{0.595} & \oodfirst{2.990} \\
        \cmidrule(lr){2-8}
        & \multirow{3}{*}{CoCoEmo} & \multirow{3}{*}{IndexTTS2} & ID-SS & 0.017 & 0.333 & \idssfirst{0.703} & 2.294 \\
        & & & ID-CS & 0.007 & 0.378 & \idcsfirst{0.699} & 2.344 \\
        & & & OOD & 0.026 & 0.356 & \oodfirst{0.688} & 2.314 \\
        \cmidrule(lr){2-8}
        & \multirow{3}{*}{EmoSteer-TTS} & \multirow{3}{*}{F5-TTS} & ID-SS & 0.013 & 0.139 & 0.465 & 2.309 \\
        & & & ID-CS & 0.013 & 0.139 & 0.465 & 2.250 \\
        & & & OOD & 0.013 & 0.139 & 0.465 & 2.243 \\
        \cmidrule(lr){2-8}
        & \multirow{3}{*}{EmoSteer-TTS} & \multirow{3}{*}{CosyVoice~2} & ID-SS & -0.009 & 0.133 & 0.362 & \idsssecond{2.575} \\
        & & & ID-CS & -0.009 & 0.133 & 0.362 & \idcsthird{2.655} \\
        & & & OOD & -0.009 & 0.133 & 0.362 & 2.542 \\
        \cmidrule(lr){2-8}
        & \multirow{3}{*}{EmoSteer-TTS} & \multirow{3}{*}{IndexTTS2} & ID-SS & -0.005 & 0.400 & \idsssecond{0.683} & 2.355 \\
        & & & ID-CS & -0.005 & 0.400 & \idcssecond{0.683} & 2.335 \\
        & & & OOD & -0.005 & 0.400 & \oodsecond{0.683} & 2.313 \\
        \midrule
        \multirow{9}{*}{\textbf{SEmoEdit}} & \multirow{3}{*}{\textbf{Ours}} & \multirow{3}{*}{F5-TTS} & ID-SS & \idsssecond{0.175} & \idsssecond{0.039} & 0.307 & 1.967 \\
        & & & ID-CS & \idcssecond{0.146} & \idcsthird{0.011} & 0.342 & 2.094 \\
        & & & OOD & \oodsecond{0.114} & \oodsecond{0.011} & 0.271 & 2.083 \\
        \cmidrule(lr){2-8}
        & \multirow{3}{*}{\textbf{Ours}} & \multirow{3}{*}{CosyVoice~2} & ID-SS & \idssthird{0.172} & \idssfirst{0.017} & 0.362 & \idssthird{2.499} \\
        & & & ID-CS & \idcsthird{0.138} & \idcsfirst{0.000} & 0.371 & \idcssecond{2.700} \\
        & & & OOD & \oodthird{0.099} & \oodfirst{0.006} & 0.309 & \oodsecond{2.870} \\
        \cmidrule(lr){2-8}
        & \multirow{3}{*}{\textbf{Ours}} & \multirow{3}{*}{IndexTTS2} & ID-SS & \idssfirst{0.192} & 0.411 & 0.336 & 2.355 \\
        & & & ID-CS & \idcsfirst{0.206} & 0.400 & 0.344 & 2.289 \\
        & & & OOD & \oodfirst{0.190} & 0.428 & 0.282 & \oodthird{2.687} \\
        \bottomrule
    \end{tabular}
\end{table*}

\section{Experimental Results}
\label{sec:experimental_results}

\subsection{Complete In- and Out-of-Distribution Results}
\label{app:complete_benchmark_results}

Tables~\ref{tab:complete_replacement_results}--\ref{tab:complete_intensity_results} report all objective metrics for ID-SS (same dataset and speaker), ID-CS (same dataset, cross speaker), and OOD (cross dataset and speaker). Unsupported methods are omitted, and intensity preservation and quality are measured at $\alpha=1$. Rankings are computed independently within each setting. Pink, green, and blue cells denote ID-SS, ID-CS, and OOD, respectively; dark, medium, and light shades mark first, second, and third place, with displayed ties sharing a rank.

\end{document}